\documentclass{article}
\usepackage[english]{babel}
\usepackage[letterpaper,top=2cm,bottom=2cm,left=3cm,right=3cm,marginparwidth=1.75cm]{geometry}

\usepackage{amsmath}
\usepackage{amsfonts}
\usepackage{mathtools}
\usepackage{mathabx}
\usepackage{graphicx}
\usepackage[colorlinks=true, allcolors=blue]{hyperref}
\usepackage{authblk}
\usepackage{booktabs}
\usepackage[numbers]{natbib}
\setcitestyle{square}

\title{Generative Model for Constructing Reaction Path from Initial to Final States}
\author[1]{Akihide Hayashi \thanks{ahayashi@preferred.jp}}
\author[1]{So Takamoto}
\author[2,3]{Ju Li}
\author[1]{Yuta Tsuboi}
\author[1]{Daisuke Okanohara}

\affil[1]{Preferred Networks, Inc., Tokyo, Japan.}
\affil[2]{Department of Materials Science and Engineering,
Massachusetts Institute of Technology, Cambridge, MA 02139, USA}
\affil[3]{Department of Nuclear Science and Engineering,
Massachusetts Institute of Technology, Cambridge, MA 02139, USA}

\begin{document}
\maketitle

\begin{abstract}
Mapping the chemical reaction pathways and their corresponding activation barriers is a significant challenge in molecular simulation. Given the inherent complexities of 3D atomic geometries, even generating an initial guess of these paths can be difficult for humans. This paper presents an innovative approach that utilizes neural networks to generate initial guesses for reaction pathways based on the initial state and learning from a database of low-energy transition paths.
The proposed method is initiated by inputting the coordinates of the initial state, followed by progressive alterations to its structure. This iterative process culminates in the generation of the guess reaction path and the coordinates of the final state.  The method does not require one-the-fly computation of the actual potential energy surface, and is therefore fast-acting. The application of this geometry-based method extends to complex reaction pathways illustrated by organic reactions.
Training was executed on the Transition1x dataset of organic reaction pathways. The results revealed the generation of reactions that bore substantial similarities with the test set of chemical reaction paths. The method's flexibility allows for reactions to be generated either to conform to predetermined conditions or in a randomized manner.
\end{abstract}

\section{Introduction}
The enhanced comprehension of chemical reactions using computational methods is continually advancing. Notably, the intersection of machine learning and computational chemistry has recently demonstrated significant potential for the exploring materials based on atomistic energetics. Recent advances in machine learning have accelerated research in computational chemistry. The advent of machine learning potentials has significantly sped up molecular dynamics. However, because chemical reactions are intrinsically rare, the acceleration provided by machine learning potentials alone is insufficient to tracking chemical reactions within feasible timeframes. Therefore, standard sampling techniques, such as conventional molecular dynamics methods or Monte Carlo methods, remain inadequate, even with improved potential speeds. Further development of the sampling techniques is required.

Significant developments in sampling techniques have occurred in recent years, with the advent of generative models equivariant to translation, rotation, and permutation. In particular, focusing on machine learning-based sampling methods, the current trend involves transforming simple distributions, such as Gaussian distributions, into complex distributions that the data should follow. Specifically, methods such as normalizing flows\cite{Rezende2015-dj}, diffusion models\cite{Ho2020-an}, and flow matching\cite{Lipman2022-jt} have been extensively studied for the sampling of molecular structures. A common feature of these methods is the iterative transformation of a simple distribution, such as a Gaussian distribution, into a target distribution.

For example, the $ E \left( n \right) $ equivariant normalizing flow \cite{Garcia_Satorras2021-dw} learns to generate actual molecules by sampling the coordinates of each atom constituting a molecule from a Gaussian distribution through a normalizing flow. Similarly, GeoDiff \cite{Corso2022-xc} learns the direction to generate molecular structures by sampling the coordinates of each atom from a Gaussian distribution but uses a diffusion model to define this direction. DiffDock \cite{Corso2022-xc} and Torsional Diffusion \cite{Jing2022-cq} introduced a coordinate system employing dihedral angles and translational degrees of freedom within the molecule and applied diffusion models within that framework. Equivariant flow matching \cite{Klein2023-bj} transforms the coordinates of each atom from a Gaussian distribution and definines the direction of atom movement using flow matching. Distributional Graphormer (DiG) \cite{Zheng2023-bv} determines the movement direction using a diffusion model and learning with various coordinates specific to each system. CDVAE\cite{Xie_undated-or} generates bulk systems using a VAE-conditioned flow. By training lightweight energy predictors from the VAE\cite{Chen2024-ds} features, it allows for the estimation of the energy of the structures before generation, thus saving time in what is typically a time-consuming process.

In addition to generating stable structures, generating reaction pathways (RPs) are in significat demand. RPs and transition states (TSs) provide crucial information regarding chemical reactions. TS is the highest energy point of the minimum energy path (MEP). The height of the transition state is a critical parameter that determines the rate of chemical reactions.

In recent years, methods that can sample RPs have also been proposed in addition to models that generate stable structures. Diffusion Methods for Generating Transition Paths \cite{Triplett2023-gg} discretize the reaction pathway and use a diffusion model in a space defined by the product of the number of degrees of freedom of the structure and the number of discrete image points. In contrast, the Boltzmann Generator \cite{Noe2019-uj} and DiG directly interpolate between two points on a Gaussian distribution, generating structures from each point to obtain a pathway connecting different basins. Notably, DiG is trained on various systems using  Graphormer \cite{Ying2021-sh}, suggesting its potential for general application. These methods can smoothly interpolate between two basins. However, because they do not use MEP information during training, it is uncertain whether the generated pathways are close to the MEP.

The lattice-free extension of Kinetic Monte Carlo (KMC) \cite{Sickafus2007-rs} can be considered a future reaction pathway generation application. For instance, recent attempts have been made to combine KMC with reinforcement learning \cite{Tang2023-ot}.

To accelerate the KMC by generating reaction pathways, it is essential to quickly enumerate the initial states (ISs), final states (FSs), and activation barriers. Reaction generation methods have been developed to handle small molecules on solids or solid surfaces. However, models capable of generating chemical reactions in organic chemistry in a continuous space have not been proposed. Therefore, to the best of our knowledge, the most promising method for application in organic reactions and the potential for accelerating reaction simulations is temperature-accelerated dynamics (TAD) \cite{Sorensen2000-gs}, which samples high-temperature MD. One reason for this is that reactions in organic compounds involve curvilinear reaction pathways in which the degrees of freedom of various atoms are interdependent. This complex degree of freedom makes it challenging to handle organic compounds in a $3N$-dimensional space.

Because generative models are generally inaccurate, providing precise MEPs or TSs where the force is almost zero under an actual potential is challenging. However, if an approximate shape of the reaction pathway can be provided, methods such as CI-NEB \cite{Henkelman2000-kb} can be used to optimize the reaction pathway or models that predict the activation energy from the approximate shape of the reaction pathway can be employed to estimate the activation barrier. Therefore, a method to rapidly generate approximate reaction pathways is required to accelerate the simulation of organic compounds.

In this study, we developed a model capable of generating appropriate organic reaction pathways from instructions or randomly arbitrary without instructions. This breakthrough approach effectively handles the complex degrees of freedom associated with organic molecules. The proposed method uses the IS structure and any reaction type as inputs. It gradually modifies the structure of the IS along the RP to obtain the approximate RP and the FS simultaneously.

To achieve this, we introduced two fields: transformation guidance and denoising. This method produces a reaction pathway connecting the IS and FS by following these fields to modify the structure. This approach is rapid and can directly learn RPs from reaction pathway datasets. In addition, it is highly versatile, as demonstrated by training a generalized model on transition1x \cite{Schreiner2022-eb} data. Moreover, the model can generate reactions for molecules with more atoms than those in the transition1x dataset.

\section{Method}

\subsection{Training target}
The number of atoms was $\mathit{N}$. Let $\mathbf{x} \in \mathbb{R}^{3\mathit{N}}$ denote the coordinates of each atom.
In general, there are numerous reaction pathways exist, among which the $\mathit{i}$th reaction pathway is denoted by $\mathbf{x}_{\mathrm{RP},\mathit{i}}(\mathit{s})$. Here, $s$ is a parameter satisfying,
\begin{equation}
    \left\lVert\frac{\mathrm{d}\mathbf{x}_{\mathrm{RP},\mathit{i}}}{\mathrm{d}\mathit{s}}\right\rVert=1.
    \label{eq:reaction-path-parameter-norm}
\end{equation}
In this context, the minimum value of $s$ is 0, and its maximum value is the length of $\mathbf{x}_{\mathrm{RP},\mathit{i}}$. Furthermore, let the length of the reaction pathway be $\mathit{L}_\mathit{i}$.
Consider the foot of the perpendicular drop from any coordinate $\mathbf{x}$ to $\mathbf{x}_{\mathrm{RP},\mathit{i}}$. At the foot of the perpendicular, the parameter $s$ coincides with $\hat{s}_\mathit{i}(\mathbf{x})$ defined in,
\begin{equation}
    \hat{s}_\mathit{i}\left(\mathbf{x}\right)
    =
    \underset{s}{\mathrm{argmin}}
        \left(
            \left\lVert
                 \mathbf{x}_{\mathrm{RP},\mathit{i}}\left(\mathit{s}\right)
                 -
                \mathbf{x}
            \right\rVert^2
        \right). \label{eq:foot-s}
\end{equation}
In the proposed method, two types of fields are defined. The first field is a transformation guidance field. The transformation guidance field is the tangent vector of the RP, pointing from the IS to the FS, at the foot of the vector drop perpendicular to the RP that should be followed. The transformation guidance field is defined as follows:
\begin{equation}
    \mathbf{t}_{\mathrm{t},\mathit{i}}\left(\mathbf{x}\right)
    \coloneqq
    \begin{cases}
        \left.
            \frac{
                \mathrm{d} \mathbf{x}_{\mathrm{RP},\mathit{i}} \left(\mathit{s}\right)
            }{
                \mathrm{d\mathit{s}}
            }    
        \right|
        _{\mathit{s}=\hat{s}_\mathit{i}\left(\mathbf{x}\right)
        } & \left(0<s<L_i\right)
        \\
        \mathbf{0} & \left(\mathrm{other}\right)
    
    \end{cases}. \label{eq:tt}
\end{equation}
The second field is the denoising field. The denoising field is a perpendicular vector pointing from $\mathbf{x}$ to the RP and drops to the RP that should be followed. The denoising field is defined as follows:
\begin{equation}
    \mathbf{t}_{\mathrm{d},\mathit{i}}\left(\mathbf{x}\right)
    \coloneqq
    \mathbf{x}_{\mathrm{RP},\mathit{i}}\left(\hat{s}_\mathit{i}\left(\mathbf{x}\right)\right)-\mathbf{x}.
    \label{eq:td}
\end{equation}
The machine learning model learns $\mathbf{t}_{\mathrm{t},\mathit{i}}(\mathbf{x})$ and $\mathbf{t}_{\mathrm{d},\mathit{i}}(\mathbf{x})$. It does not have information about the index of the training data. Instead, it receives a condition vector given by the feature $\mathbf{c}$ as input. Let $\mathbf{y}_{\mathrm{t},\mathit{i}}(\mathbf{x},\mathbf{c})$ be the approximation of the training data $\mathbf{t}_{\mathrm{t},\mathit{i}}(\mathbf{x})$ by the machine learning model. Similarly, let $\mathbf{y}_{\mathrm{d}}(\mathbf{x},\mathbf{c})$ be an approximation of the training data $\mathbf{t}_{\mathrm{d},\mathit{i}}(\mathbf{x})$ using the machine learning model.
$\mathbf{t}_{\mathrm{t},\mathit{i}}(\mathbf{x})$ is the derivative of the RP for parameter. Therefore, we start with IS and integrate it into $\mathbf{t}_{\mathrm{t},\mathit{i}}(\mathbf{x})$, as shown in,
\begin{equation}
    \mathrm{d}\mathbf{x}
    =
    \mathbf{t}_{\mathrm{t},\mathit{i}}\left(\mathbf{x}\right)
    \mathrm{d}s,
    \label{eq:ideal-differential-equation-t}
\end{equation}
allows us to obtain the RP.
Therefore, one is expected to generate the RP by learning the pathway from the IS to FS. However, if one attempts to generate an RP using only $\mathbf{y}_{\mathrm{t},\mathit{i}}$, the path may deviate from the actual RP because of inference or approximation errors. The field at positions far from the RP is not well learned, and there is no physical meaning to moving along a tangent to the RP at such positions. Consequently, once the pathway deviates from the RP, it diverges.

To address this problem, methods such as score matching and diffusion models learn where to move, even in the vicinity where data appear, to return to the region where the data occurs. Similarly, in learning RPs, if a structure deviates from the actual RP, it is necessary to return and correct it in the direction of the RP.

For this purpose, generation is performed using a linear combination of the transformation guidance and denoising fields, as shown in
\begin{equation}
    \mathrm{d}\mathbf{x}
    =
    \mathbf{y}_{\mathrm{t}}\left(\mathbf{x},\mathbf{c}\right)\mathrm{d}\mathit{s}+\alpha\mathbf{y}_{\mathrm{d}}\left(\mathbf{x},\mathbf{c}\right).
    \label{eq:denoising-differential-equation}
\end{equation}
In addition, unlike general diffusion models, $s$ is a parameter corresponding to the length of the RP, and its maximum value is unknown during generation. Therefore, to determine whether we are partway along the RP, we define $t_\mathrm{f}$ as a variable. $t_\mathrm{f}$ is an integer scalar output that takes the value of zero or one. Additionally, let $\mathbf{y}_{\mathrm{f}}(\mathbf{x}, \mathbf{c})$ be the approximation of $t_{\mathrm{f},\mathit{i}}$ using a machine learning model. $\mathbf{y}_{\mathrm{f}}(\mathbf{x}, \mathbf{c})$ outputs two values, and the generation stops based on which value is larger.

\begin{figure}
\centering
\includegraphics[width=0.9\textwidth]{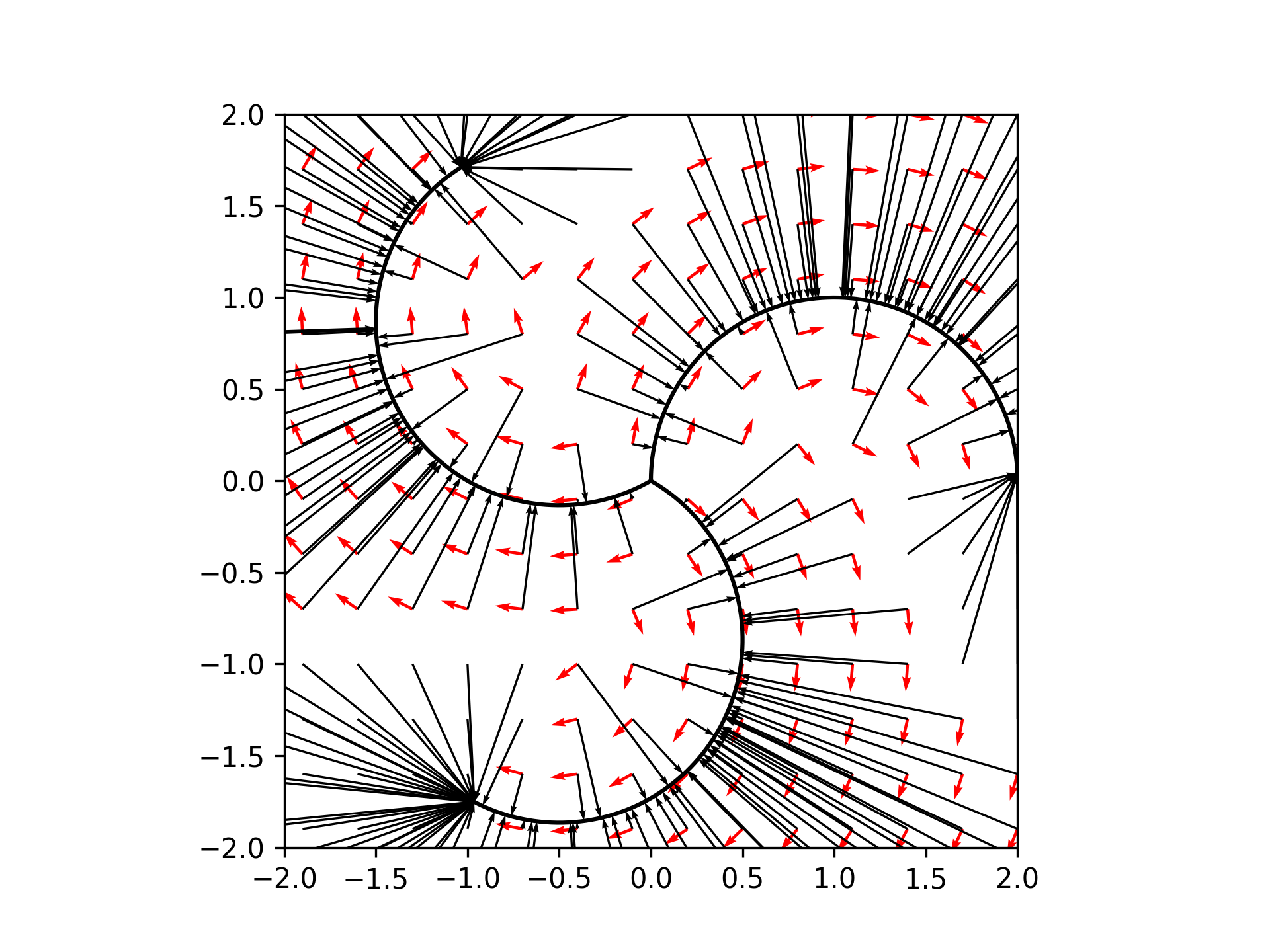}
\caption{\label{fig:flow} Model image of the learned flow. Consider the center of the figure as the IS. The three curves emanating from the IS are considered optimized RPs. The FS is the endpoint of the RPs on the side opposite to the IS. The black arrows extending from each point in the figure represent the denoising field ($\mathbf{t}_\mathrm{d}$), and the red arrows represent the transformation guidance field ($\mathbf{t}_\mathrm{t}$).}
\end{figure}

For clarity, Fig.~\ref{fig:flow} shows a model of the RP as a curve in a 2D space. In Fig.~\ref{fig:flow}, the center of the figure represents the IS. The three curves emanating from the IS each represent different RPs as indicated by the expressions $\mathbf{x}_{\mathrm{RP},\mathit{i}}, i \in [1, 3]$.

When such RPs exist, the figure shows the field used as the training data when the RP with the shortest denoising field is selected for learning at each coordinate $\mathbf{x}$. In Fig.~\ref{fig:flow}, the red arrows denote the transformation guidance field, and the thin black arrows denote the denoising field.
\subsection{Related Work}
Flow matching, diffusion models, and reinforcement learning significantly influenced the model used in this study with many modifications. The transformation guidance and denoising fields can be interpreted within the context of the flow matching and diffusion models. Furthermore, the differences in problem settings between this approach and imitation learning are discussed.

\subsubsection{Relationship between the Transformation Guidance Field and Existing Generative Models}
The transformation guidance field can be interpreted as a type of flow matching. In general diffusion models or flow matching, a "time" concept connects a simple distribution with the generated distribution smoothly. For example, at \( t = 0 \), the molecules follow a Gaussian distribution, and at \( t = 1 \), they follow a a Boltzmann distribution. However, the transformation guidance field contains elements distinct from those in general diffusion models or flow matching. In the transformation guidance field, a distribution localized near the IS was used instead of a simple distribution, and a distribution localized around FS was used as the generated distribution. 

In flow matching or diffusion models, the pathway during generation is not crucial; only the distribution of the FS is essential. In contrast, the transformation guidance field considers the necessary pathways, and all the structures obtained during the generation process are used to construct the RP. Furthermore, in the present problem setting, a time-independent vector field is learned, which starkly contrasts with diffusion models or flow matching, which learn time-varying vector fields. The pathway that transitions from IS along the learned vector field becomes the RP.

Because the vector field does not depend on time, unlike general flow matching or diffusion models, the number of steps required to complete the generation is not uniquely determined. Therefore, we  predict the stopping condition using Eq.~\ref{eq:fin}.

\subsubsection{Relationship between the Denoising Field and Existing Generative Models}

The denoising field is related to denoising score matching\cite{Vincent2011-xz} in a scenario in which every point on the RP represents the data distribution. Suppose the RP can be approximated as a discrete set of points and the nearest point on the RP to a given point $\mathbf{x}$ is unique. In that case, it can be proven that the denoising and Newton steps to maximize the log-likelihood for the perturbed distribution in denoising score matching are equivalent. Let \(\mathbf{x}_{\mathrm{RP},i,s}\) denote the \(s\)-th discretized point in the \(i\)-th RP. In denoising score matching, the distribution diffused around the data points is given by
\begin{equation}
    p_i\left(\mathbf{x},\sigma\right)
    =\frac{1}{S_i}\sum_{s=1}^{S_i}\mathcal{N}\left(\mathbf{x};\mathbf{x}_{\mathrm{RP},i,s},\sigma\right) \label{eq:noised-data-distribution}
\end{equation}
where \(\mathcal{N}\) is the probability density function of a normal distribution represented by
\begin{equation}
\mathcal{N}(\mathbf{y} ; \mathbf{x}, \sigma) \equiv \frac{1}{\sqrt{2 \pi}^{3 N} \sigma^{3 N}} \exp \left(-\frac{\|\mathbf{x}-\mathbf{y}\|^2}{2 \sigma^2}\right) .\label{eq:normal-distribution}
\end{equation}
The gradient of the logarithm of Eq. ~\ref{eq:noised-data-distribution} with respect to \(\mathbf{x}\) is given by
\begin{equation}
    \nabla \log{p_i\left(\mathbf{x},\sigma\right)}
    =
    \frac{1}{p_i\left(\mathbf{x},\sigma\right)}
    \frac{1}{S_i}\sum_{s=1}^{S_i}
    \frac{\mathbf{x}_{\mathrm{RP},i,s}-\mathbf{x}}{\sigma^2}\mathcal{N}\left(\mathbf{x};\mathbf{x}_{\mathrm{RP},i,s},\sigma\right).
    \label{eq:noised-data-distribution-gradient}
\end{equation}
Furthermore, the second derivative of the logarithm of Eq.~\ref{eq:noised-data-distribution} with respect to \(\mathbf{x}\) is given by,
\begin{equation}
\begin{aligned}
    &\nabla\nabla^\top \log{p_i\left(\mathbf{x},\sigma\right)}
    \\&=
    \frac{1}{p_i\left(\mathbf{x},\sigma\right)}
    \frac{1}{S_i}\sum_{s=1}^{S_i}
    \left(
        \left(
            \frac{\mathbf{x}_{\mathrm{RP},i,s}-\mathbf{x}}{\sigma^2}
        \right)
        \left(
            \frac{\mathbf{x}_{\mathrm{RP},i,s}-\mathbf{x}}{\sigma^2}
        \right)^\top
        -
        \left(
            \frac{\mathbf{I}_{3N}}{\sigma^2}
        \right)
    \right)
    \mathcal{N}\left(\mathbf{x};\mathbf{x}_{\mathrm{RP},i,s},\sigma\right)
    \\&-
    \nabla \log{p_i\left(\mathbf{x},\sigma\right)}
    \nabla^\top \log{p_i\left(\mathbf{x},\sigma\right)}. \label{eq:noised-data-distribution-hessian}
    \end{aligned}
\end{equation}
Here, we introduce the Soft-Nearest function represented in the form of ,
\begin{equation}
\mathrm { SN }_i[f, \mathbf{x},\sigma]:=\frac{1}{p_i(\mathbf{x}, \sigma)} \frac{1}{S_i} \sum_{s=1}^{s_i} f\left(\mathbf{x}, \mathbf{x}_{\mathrm{RP}, \mathrm{i}, \mathrm{s}}\right) \mathcal{N}\left(\mathbf{x} ; \mathbf{x}_{\mathrm{RP}, \mathrm{i}, \mathrm{s}}, \sigma\right). \label{eq:soft-nearest}
\end{equation}
Now, $\hat{s}_\mathit{i}\left(\mathbf{x}\right)$ is introduced as:
\begin{equation}
    \hat{s}_i(\mathbf{x})=\underset{s}{\mathrm{argmin}}
    \left(
            \left\lVert
                 \mathbf{x}_{\mathrm{RP},\mathit{i},\mathit{s}}
                 -
                \mathbf{x}
            \right\rVert^2
        \right).
        \label{eq:foot-s-discrete}
\end{equation}
When $\forall \mathit{s}\neq\hat{\mathit{s}}_\mathit{i}(\mathbf{x}),~
\left\lVert
                 \mathbf{x}_{\mathrm{RP},\mathit{i},\hat{\mathit{s}}_\mathit{i}(\mathbf{x})}
                 -
                \mathbf{x}
    \right\rVert
    <
    \left\lVert
                 \mathbf{x}_{\mathrm{RP},\mathit{i},\mathit{s}}
                 -
                \mathbf{x}
    \right\rVert$, in the limit as \(\sigma \rightarrow 0\), the Soft-Nearest function satisfies,
\begin{equation}
    \mathrm{SN}_i\left[f,\mathbf{x},\sigma\right]
    =
    f\left(\mathbf{x},\mathbf{x}_{\mathrm{RP},i,\hat{s}_i(\mathbf{x})}\right)
    +
O\left(\exp\left(-\frac{1}{\sigma}\right)\right)
    \label{eq:soft-nearest-convergence}
\end{equation}
converges to the value at the point on the RP closest to \(\mathbf{x}\). 
Because of this property, the denoising field coincides with the \(\sigma \rightarrow 0\) limit of the following equation.
\begin{equation}
    \mathbf{t}_{\mathrm{d},\mathit{i}}\left(\mathbf{x},\sigma\right)
    =
    \mathrm{SN}_i\left[\mathbf{x}_{\mathrm{RP},i,s}-\mathbf{x},\mathbf{x},\sigma\right],
    \label{eq:approximated-denoising-field}
\end{equation}
\begin{equation}
    \lim_{\sigma\rightarrow0}\mathbf{t}_{\mathrm{d},\mathit{i}}\left(\mathbf{x},\sigma\right)
    =
    \mathbf{t}_{\mathrm{d},\mathit{i}}\left(\mathbf{x}\right) \label{eq:noised-denoising-field}.
\end{equation}
To investigate how far the Newton step and $\mathbf{t}_{\mathrm{d},\mathit{i}}\left(\mathbf{x},\sigma\right)$ are for any $\sigma$, we define the difference between the Hessian of the log-probability density function applied to $\mathbf{t}_{\mathrm{d},\mathit{i}}\left(\mathbf{x},\sigma\right)$ and the gradient of the log-probability density function as shown in,
\begin{equation}
    R_{i}\left(\mathbf{x},\sigma\right)
    \coloneqq
    \nabla\nabla^\top \log{p_i\left(\mathbf{x},\sigma\right)}\mathbf{t}_{\mathrm{d},\mathit{i}}\left(\mathbf{x},\sigma\right)
    +
    \nabla \log{p_i\left(\mathbf{x},\sigma\right)}.
    \label{eq:remaining}
\end{equation}
 This equation can be expanded as,
\begin{equation}
\begin{aligned}
    &R_i(\mathbf{x}, \sigma)
    \\&=
    \frac{1}{\sigma^4}\left(
        \mathrm { SN }_i\left[\left(\mathbf{x}
        -\mathbf{x}_{\mathrm{RP}, \mathrm{i}, \mathrm{s}}\right)\left(\mathbf{x}-\mathbf{x}_{\mathrm{RP}, \mathrm{i}, \mathrm{s}}\right)^{\top}\right]-\mathrm { SN }_i\left[\left(\mathbf{x}-\mathbf{x}_{\mathrm{RP}, \mathrm{i}, \mathrm{s}}\right)\right] \mathrm { SN }_i\left[\left(\mathbf{x}-\mathbf{x}_{\mathrm{RP}, \mathrm{i}, \mathrm{s}}\right)^{\top}\right]\right) \mathrm { SN }_i\left[\left(\mathbf{x}-\mathbf{x}_{\mathrm{RP}, \mathrm{i}, \mathrm{s}}\right)\right]. \label{eq:remaining-expand}
\end{aligned}
\end{equation}
In Eq.~\ref{eq:remaining-expand}, the contents of the parentheses asymptotically approach 0 as $\sigma \rightarrow 0$ with an order of $O\left(\exp(-\frac{1}{\sigma})\right)$. Therefore, $R_i(\mathbf{x}, \sigma)$ asymptotically approaches 0 as $\sigma \rightarrow 0$, indicating that $\mathbf{t}_{\mathrm{d},\mathit{i}}\left(\mathbf{x}\right)$ is a Newton step in the log-probability density function as $\sigma \rightarrow 0$. Consequently, for the denoising field term, in many cases, when $\alpha = 1$, it will yields values close to the Newton step.

Instead of using a denoising field, we can consider using an orthogonalized potential force with respect to the transformation guidance field. However, the orthogonalized force has an inverse dimension of distance, which differs from the transformation guidance field and the coordinates. Therefore, to combine it linearly with the transformation guidance field, it is necessary to multiply it by a constant with the appropriate magnitude and dimension, which can vary depending on the system, making the adjustment challenging. In this respect, the denoising field is more convenient because it has length dimensions.

\subsubsection{Relationship with Imitation Learning}
Reaction pathway prediction can also be considered a sequential decision-making problem, in which the task is to predict the coordinates along the MEP at each time step. One approach to this problem is imitation learning (behavior cloning), which involves supervised learning from the trajectories of correct actions.  \citet{ross2011reduction} proposes algorithms that address the challenge of the difference between the correct trajectory and the inference-time trajectory, which can lead to unobserved states during training and result in failed predictions. The denoising field proposed in this study learns the direction perpendicular to the correct trajectory. It serves as a similar solution by recovering to the correct trajectory. However, by removing noise, our denoising field learns the score function, which is the gradient of the log-likelihood. Because the gradient of the log-likelihood includes $p_i\left(\mathbf{x},\sigma\right)$ in the denominator, as shown in Eq.~\ref{eq:noised-data-distribution-gradient}, the score can be large even in low-probability regions, allowing it to return to high-probability regions.
Although they and their subsequent research  utilized a trained model (policy) to generate the training data~\cite{ross2011reduction,he2012imitation}, we generated the training data by adding random noise to the correct points in our experiments.  We believe that it is worth exploring other approaches to training data generation..

However, because \citet{ross2011reduction} and most reinforcement learning settings assume discrete time steps, the direction from a divergent point to the next step on the correct trajectory can be trivially defined. The models are trained to predict that direction directly. In contrast, in this study, because the reaction pathway must be obtained by integrating $\mathbf{t}_{\mathrm{t},\mathit{i}}\left(\mathbf{x}\right)$, the definition of the next step in continuous time is not trivial. Therefore, in this study, the problem was modeled using two fields: transformation guidance field and denoising.

\section{Training}

\subsection{Notation}
\(\bigotimes\) denotes the tensor product in e3nn \cite{Geiger2022-om}. \(\bigoplus\) represents concatenation. \(\bigodot\) denotes the operation that takes the product of two features and sums them in the feature direction.

Symbols with a single subscript, such as \(Z_i\), indicate the notation for the \(i\)-th node (the \(i\)-th atom). Symbols with two subscripts, such as \(c_{ij}\), represent the edges between the \(i\)-th and \(j\)-th nodes. Additionally, the values in bold, such as \(\mathbf{r}\), are $E(3)$-equivariant quantities. \(\hat{\mathbf{r}}\) represents the normalized \(\mathbf{r}\), and \(\lVert \mathbf{r} \rVert\) represents the norm of \(\mathbf{r}\).

All variables are written in e3nn notation according to the transformation rules to which they belong. A particularly important note is that the \(0e\) components are $E(3)$-invariant scalar components, and the \(1o\) components are $E(3)$-equivariant vector components. In addition, \(128 \times 0e\) indicates a feature consisting of 128 $E(3)$-invariant scalar components.
"FNN" indicates the operation where a fully connected neural network is applied to the \(0e\) components of each atom. The SiLU activation function was used \cite{Hendrycks2016-hs}. For transformations in which the number of inputs and outputs were the same, a ResBlock \cite{He2016-ij} was employed.

\subsection{Neural network architecture}\label{sec:whole-model}
\begin{figure}
    \centering
    \includegraphics[width=1\linewidth]{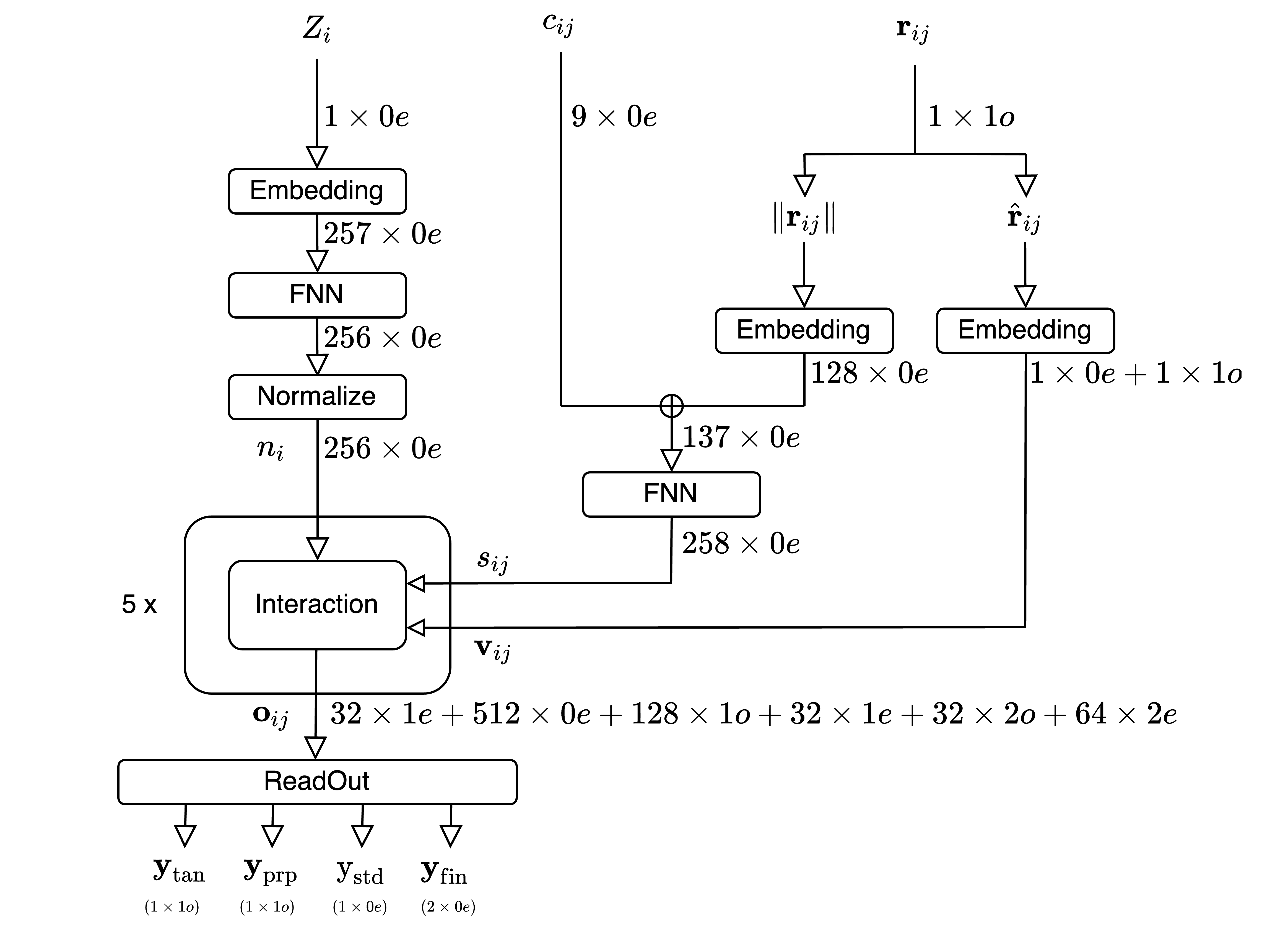}
    \caption{The whole model architecture.}
    \label{fig:model-whole}
\end{figure}

Let \(i\) and \(j\) be atom indices. The overall structure of the model is as follows: The formal inputs to the model were as follows:
\begin{equation}
    f_{\mathrm{cart}}:\{\mathbf{x}_{i}, \mathbf{x}_{\mathrm{IS},i}, Z_i, c_{ij}\} \longrightarrow \{\mathbf{y}_{\mathrm{t},i},\mathbf{y}_{\mathrm{d},i},\mathit{y}_{\mathrm{std},i},\mathit{y}_{\mathrm{f}}\ \}. \label{eq:model-input-raw}
\end{equation}
where, \(\mathbf{x}\) represents the coordinates of the current structure, and \(\mathbf{x}_{\mathrm{IS}}\) represents the coordinates of the IS. \(Z_i\) is the atomic number. In addition, \(c_{ij}\) is a feature vector specifying the generated reaction. In this experiment, \(c\) only handled the edge features.
Only relative coordinates were used to guarantee translational and rotational equivariance. The inputs and outputs obtained using the relative coordinates are shown in Fig.~\ref{fig:model-whole} and,
\begin{equation}
    f_{\mathrm{rel}}:\{\mathbf{r}_{ij},Z_i, c_{ij}\}\longrightarrow \{\mathbf{y}_{\mathrm{t},i},\mathbf{y}_{\mathrm{d},i},\mathit{y}_{\mathrm{std},i},\mathit{y}_{\mathrm{f}}\}. \label{eq:model-inout}
\end{equation}
Here, \(\mathbf{r}\) represents the relative coordinates between the atoms. There are three types of relative coordinates: 1) the relative coordinates between atoms in the IS structure, \(\mathbf{x}_{\mathrm{IS},i} - \mathbf{x}_{\mathrm{IS},j}\), 2) the relative coordinates between atoms in the current structure, \(\mathbf{x}_{i} - \mathbf{x}_{j}\), and 3) the relative coordinates between atoms in the IS and the current structure, \(\mathbf{x}_{i} - \mathbf{x}_{\mathrm{IS},i}\). These are collectively denoted by \(\mathbf{r}_{ij}\). For \(\mathbf{x}_{i} - \mathbf{x}_{\mathrm{IS},i}\), only the edges where information flowed from \(\mathbf{x}_{\mathrm{IS},i}\) to \(\mathbf{x}_{i}\) were used.
The pairs \((i, j)\) are determined with reference to Big Bird \cite{Zaheer2020-rb}. Big Bird is a method to sparsify the attention edges in Transformers. In Big Bird, in addition to regular nodes, supernodes are prepared. Edges are connected between nodes that are close to each other, all nodes and supernodes, and randomly selected pairs of nodes. These three types of connections are called window, global, and random connections.
In this experiment, we selected and connected up to the 128 nearest neighbors among the atoms within a distance of 12 Å (windows). In addition, 32 atoms were randomly connected within a distance of 30 Å (random). We also prepared two supernodes and connected all the atoms to the supernodes (global). Because supernodes do not have coordinates, the relative coordinates of the edges between supernodes are set to \(\mathbf{r}_{ij} = \mathbf{0}\).
In Eq.~\ref{eq:model-inout}, \(c_{ij}\) is composed of a zero or one value, which indicates whether certain conditions are satisfied. The first condition denotes whether a bond is broken by the reaction (i.e., whether the atoms move far apart). The second condition indicates whether the reaction forms a bond (i.e., whether the atoms come closer together). The third condition indicates whether the dihedral angle around the bond is rotated by 105° or more because of the reaction. The third condition is set to zero if the first or the second condition is satisfied. The fourth condition indicated whether the first three conditions were used as the model input. If the fourth condition is set to 0, the first through the third conditions are set to zero; if set to 1, those conditions are used in the conditional generation; and if set to 0, they are used in the unconditional generation. The fifth condition indicates whether the edge is an edge connects the IS and the current structure for each atomic bond. The sixth condition determines whether an edge is a window. The seventh condition indicates whether the edge is a random. The eighth condition indicates whether the edge has an atom-supernode connection. The ninth condition indicates whether the edge is a supernode-supernode connection.

Fig.~\ref{fig:model-whole} shows the entire model. A one-hot vector is used for the embedding \(Z_i\). The embedding of \(Z_i\) is further transformed for each node by a neural network and then normalized and treated as a node feature with only \(0e\) components, which serves as the input to the first interaction block. The relative coordinates \(\mathbf{r}_{ij}\) are decomposed into relative distances \(\lVert \mathbf{r}_{ij} \rVert\) and normalized relative coordinates \(\hat{\mathbf{r}}_{ij}\), each of which is embedded separately. Sinusoidal embedding \cite{Vaswani2017-ku} was used to embed \(\lVert \mathbf{r}_{ij} \rVert\), and e3nn spherical harmonics \cite{Geiger2022-om} were used to embed \(\hat{\mathbf{r}}_{ij}\). Here, the \(1 \times 0e\) component is always one, and the \(1 \times 1o\) component is normalized \(\hat{\mathbf{r}}_{ij}\).

Embedding of the relative distance is a scalar edge feature that satisfies the same transformation rule ($E(3)$-invariant) as $c_{ij}$. Therefore, it is concatenated with $c_{ij}$ and treated as a edge scalar feature. After concatenation, the result transformed by the FNN is treated as the edge scalar feature $S_{ij}$. It is used as the input to the interaction block. The embedding of the normalized relative coordinates is treated as an edge vector feature $\mathbf{v}_{ij}$. It is used as the input to the interaction block.

The interactions were performed five times, during which $S_{ij}$ and $\mathbf{v}_{ij}$ were fixed. However, the node features have different values each time. Furthermore, because the tensor product of the node features and $\mathbf{v}_{ij}$ is included in the interaction block, the node features acquire higher-order tensor features for each interaction.

\subsection{Interaction}\label{sec:interaction}
\begin{figure}
    \centering
    \includegraphics[width=1.0\linewidth]{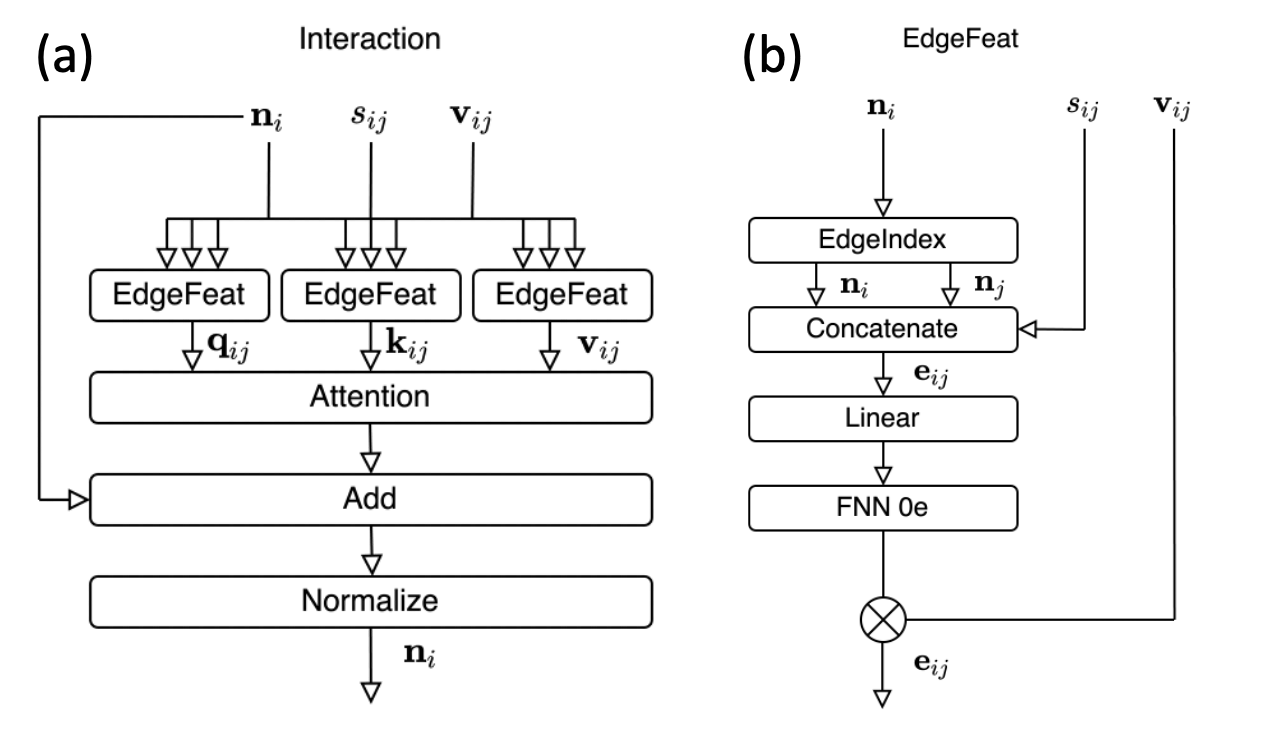}
    \caption{Diagrams of the interaction of the $E(3)$-attention network. (a) The whole diagram. (b) Details of the EdgeFeat block used in (a).}
    \label{fig:model-interaction}
\end{figure}

The interaction part of Fig.~\ref{fig:model-whole} is illustrated in Fig.~\ref{fig:model-interaction} (a).
The $E(3)$-attention is implemented. Three edge features ($\mathbf{q}_{ij}$, $\mathbf{k}_{ij}$, $\mathbf{v}_{ij}$) are generated using the three edge feature blocks introduced in Sec.~\ref{sec:edge-feat}.
Node feature $\mathbf{n}_i$ is obtained from $\mathbf{q}_{ij}$, $\mathbf{k}_{ij}$, and $\mathbf{v}_{ij}$. The following equation represents the attention:
\begin{equation}
    w_{ij}=\frac{\mathbf{q}_{ij} \cdot \mathbf{k}_{ij}}{\sqrt{d}},\label{eq:attention-w}
\end{equation}
\begin{equation}
    \mathbf{n'}_{i}=\sum_j\left({\frac{\exp( w_{ij})}{\sum_{k}\exp( w_{ik})}\mathbf{v}_{ij}}\right) ,\label{eq:attention}
\end{equation}
where the "$\cdot$" denotes the dot product, which is the sum of the products of the corresponding elements of the features. Consequently, $w_{ij}$ is $1\times0e$. $\mathbf{n'}_{i}$ obtained from Eq.~\ref{eq:attention} is added to the input $\mathbf{n}_{i}$ of the interaction block and normalized to obtain the overall output. Here, the normalization is represented by
\begin{equation}
    \frac{\mathbf{n}_{i}}{\sqrt{\mathbf{n}_{i}\cdot\mathbf{n}_{i}}}.\label{eq:normalize}
\end{equation}
By normalizing in this method, rotational equivariance is preserved.

\subsection{EdgeFeat}\label{sec:edge-feat}
In Chapter \ref{sec:interaction}, edge features are constructed using an edge feature block. An edge feature block is shown in Fig. \ref{fig:model-interaction}(b). First, the node features are expressed as $\mathbf{n}_i$ and $\mathbf{n}_j$ to represent the interactions. A single-edge tensor feature is formed by concatenating $\mathbf{n}_i$, $\mathbf{n}_j$, and $S_{ij}$. An o3 linear transformation is then performed, and an FNN was applied. A tensor product with edge vector features is computed using the result as the input to produce the output. The tensor product incorporates high-rank components into its features.

\subsection{ReadOut}
\begin{figure}
    \centering
    \includegraphics[width=1\linewidth]{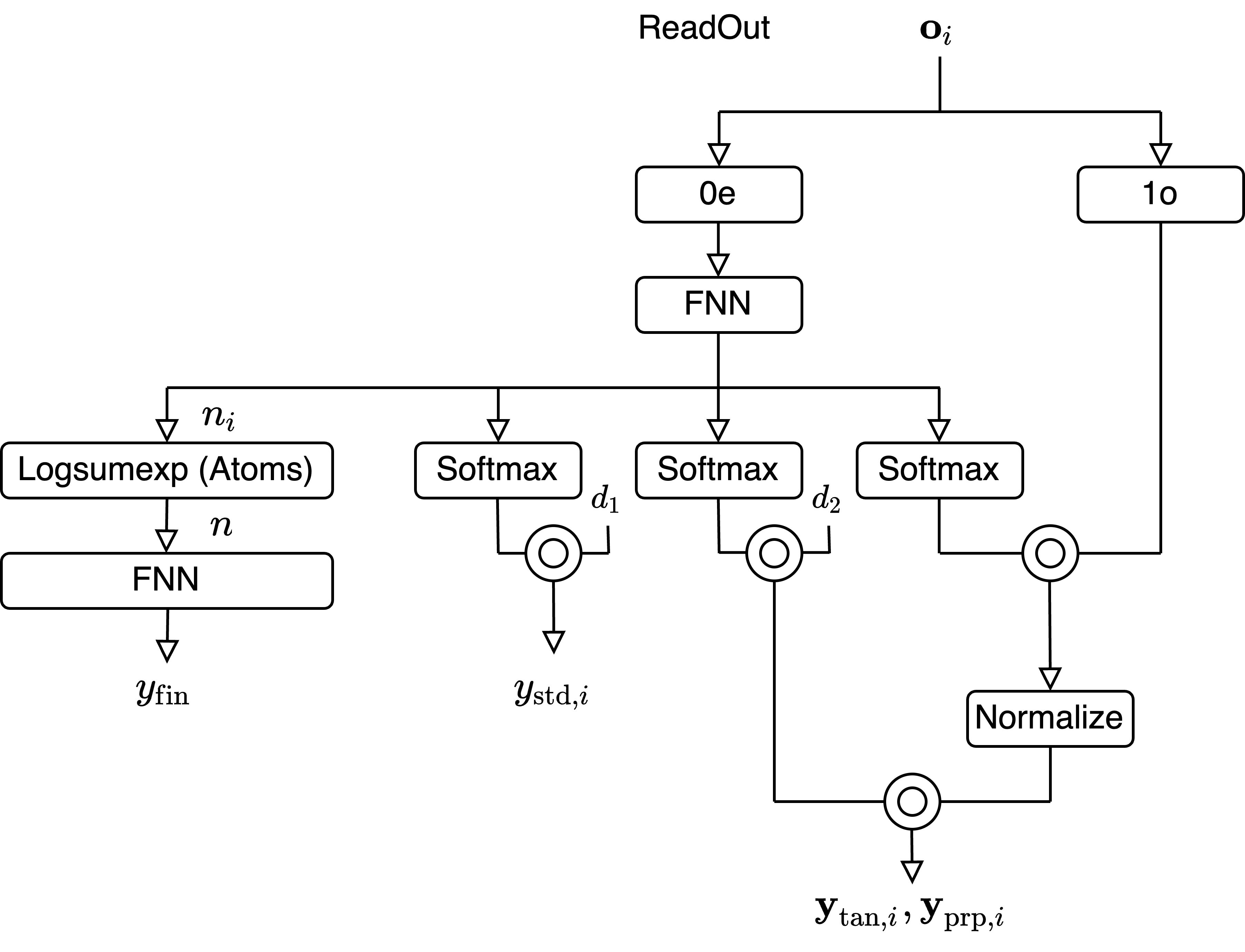}
    \caption{Model architecture of the readout part.}
    \label{fig:model-readout}
\end{figure}

Fig.~\ref{fig:model-whole} illustrates the readout section of Fig.~\ref{fig:model-readout}. There are four outputs. Output $y_{\mathrm{fin}}$ indicates the stopping condition and has a size of $2\times0e$. Output $y_{\mathrm{std},i}$ allows different prediction uncertainties for each atom and has a size of $1\times0e$. The outputs $\mathbf{y}_{\mathrm{tan},i}$ and $\mathbf{y}_{\mathrm{prp},i}$ are the direction prediction outputs. Each has a size of $1\times0o$. Using inner product decoding (Sec.~\ref{sec:inner-product-decoding}), the sizes of $\mathbf{y}_{\mathrm{tan},i}$ and $\mathbf{y}_{\mathrm{prp},i}$ are determined.

\subsection{Inner product decoding} \label{sec:inner-product-decoding}
The norms of the prediction vectors $\mathbf{y}_{\mathrm{prp}}$ and $\mathbf{y}_{\mathrm{tan}}$, and the scalar value $y_{\mathrm{std}}$ indicating the reliability of the prediction were decoded using the inner product. After applying the FNN transformation to the $0e$ component of the features, a softmax function was applied to the features. The inner product was obtained using an array of equidistant numerical values of the same dimension as the features.

The arrays used to predict the norms of $\mathbf{y}_{\mathrm{prp}}$ and $\mathbf{y}_{\mathrm{tan}}$ are evenly spaced sequences from -2$\ \textrm{\AA}$ to 2$\ \textrm{\AA}$ in steps of 0.1$\ \textrm{\AA}$. The array used to predict $y_{\mathrm{std}}$ was an evenly spaced sequence from 0.1$\ \textrm{\AA}$ to 1.0$\ \textrm{\AA}$ in steps of 0.1$\ \textrm{\AA}$.

\subsection{Dataset}
Transition1x was used. When validating the results, there are cases in which the generated results are reoptimized using the string method. To enable fast optimization in such instances, a high-speed machine learning potential known as PFP\cite{Takamoto2021-fp} is employed. Accordingly, the training data used were derived from the RP optimizations of Transition1x using PFP, which was used as the training dataset. The version of PFP used was v4.0.0 Crystal U0 mode. After reoptimization, some paths were split into multiple barriers, and 10074 paths with single barrier were generated. Among the IS and FS of these paths, molecules with an energy difference within 0.05 eV and the distance between the most-moved atoms within 0.1$\textrm{\AA}$ were considered identical. Identity determination was performed for all IS and FS. Subsequently, we extracted only those reactions that involved changes in bonding or dihedral angle rotations. Furthermore, only the reaction with the lowest activation barrier was extracted among reactions sharing the same IS and FS. This process reduces some of the pathways; however, all the pathways are duplicated twice in a round-trip manner. Consequently, 11801 reactions were identified as unique pathways. Next, the data were divided into training, validation, and testing datasets. During this process, care was taken to ensure that molecules of the same composition were grouped. Consequently, it is guaranteed that there is no data overlap between the training and validation datasets. 90\% of the total compositions were used as training data. 5\% was used as validation data, and the remaining 5\% was used as test data.

\subsection{Sampling}
The data coordinates directly used to train the neural network were sampled along the RP. We selected one RP and used a structure with noise added to the IS as the initial value. Subsequently, by moving in the direction of the RP, as shown in
\begin{equation}
    \mathrm{d}\mathbf{x}=\mathbf{t}_{\mathrm{t},\mathit{i}}\mathrm{d}t+\alpha\mathbf{t}_{\mathrm{d},\mathit{i}}+g\mathrm{d}\mathbf{w} ,\label{eq:sample-train-data}
\end{equation}
approaching RP and adding noise, we sampled the data. Here, $\mathbf{w}$ denotes the Wiener process. $g$ was uniformly sampled from the range [0.0, 0.2] each time the RP was selected. This process sampled the structures around the RP.
The resulting distribution forms a tubular shape around the RP. The distribution perpendicular to the pathway approximates the distribution obtained by scaling the data sampled from a $\chi^2$-distribution with degrees of freedom equal to the molecular degrees of freedom by a factor of $g/\sqrt{2}$.
To facilitate the learning of $t_{\mathrm{fin}}$, data with $t_{\mathrm{fin}} = 0$ and data with $t_{\mathrm{fin}} = 1$ were selected in a ratio of 1:1. To train the unconditional generation, the training for unconditional generation was conducted with a probability of 0.3. In contrast, conditional generation was conducted with a probability of 0.7.

\subsection{Training}
\begin{figure}
    \centering
    \includegraphics[width=1\linewidth]{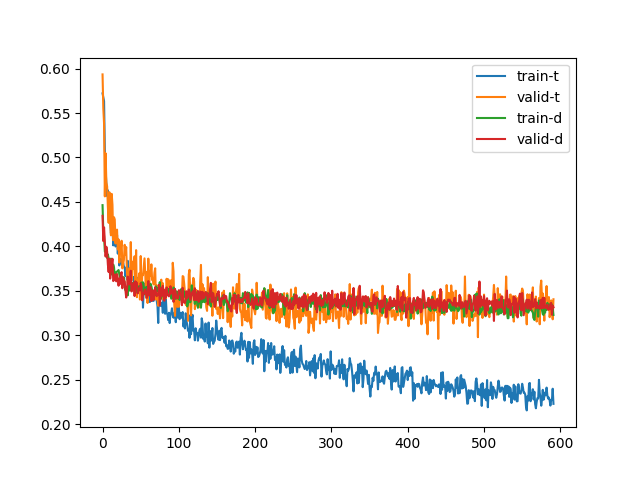}
    \caption{The mean of the norm of the difference vector between the predicted and ground truth data for both the transformation guidance field and the denoising field during the training steps. Here, train-t relates to the transformation guidance field within the training dataset, and valid-t pertains to the transformation guidance field within the validation dataset. Similarly, train-d is associated with the denoising field in the training dataset. At the same time, valid-d is connected to the denoising field in the validation dataset.}
    \label{fig:learning-curve-transition1x}
\end{figure}

The learning of $\mathbf{y}_{\mathrm{t}}$ and $\mathbf{y}_{\mathrm{d}}$ is based on the methods of score matching \cite{Ho2020-an} and flow matching \cite{Lipman2022-jt}. However, for ease of learning, the standard deviation of the output was also predicted, and the loss was defined using the standard deviation. First, a multidimensional Gaussian distribution centered on $\mathbf{t}$ is expressed as follows:
\begin{equation}
p\left(
    \mathbf{y}\left(\mathbf{x},\mathbf{c}\right)
    ;
    \mathbf{t}_{\mathit{i}}
    ,
    \mathbf{\Sigma}
    \right)
    =
    \frac{1}{\sqrt{2 \pi}^{3\mathit{N}} \sqrt{|\mathbf{\Sigma}|}}
    \exp \left(
    -\frac{1}{2}
    (\mathbf{t}_{\mathit{i}}-\mathbf{y}\left(\mathbf{x},\mathbf{c}\right))^{\top}
    \boldsymbol{\Sigma}^{-1}
    (\mathbf{t}_{\mathit{i}}-\mathbf{y}\left(\mathbf{x},\mathbf{c}\right))
    \right).
    \label{eq:gauss}
\end{equation}
Taking the negative logarithm of Eq.~\ref{eq:gauss}, we define the loss for the fields as 
\begin{equation}
\begin{aligned}
l_\mathrm{gauss}\left(
    \mathbf{y}\left(\mathbf{x}
    ,
    \mathbf{c}\right)
    ,
    \mathbf{t}_{\mathit{i}}
\right)
&=-\log p\left(
    \mathbf{y}\left(\mathbf{x}
    ,
    \mathbf{c}\right)
    ;
    \mathbf{t}_{\mathit{i}},\mathbf{\Sigma}
    \right)
\\&=
\frac{3}{2} N \log 2 \pi
+
\frac{1}{2} \log |\boldsymbol{\Sigma}|
+
\frac{1}{2}
\left(\mathbf{t}_{\mathit{i}}-\mathbf{y}\left(\mathbf{x},\mathbf{c}\right)\right)
^
{\top}
\boldsymbol{\Sigma}^{-1}
\left(\mathbf{t}_{\mathit{i}}-\mathbf{y}\left(\mathbf{x},\mathbf{c}\right)\right).
\label{eq:neg-log-gauss}
\end{aligned}
\end{equation}
For the denoising field and translation guidance fields, we define the losses as follows:
\begin{equation}
    l_\mathrm{d}=l_\mathrm{gauss}\left(
    \mathbf{y}_\mathrm{d}\left(\mathbf{x}
    ,
    \mathbf{c}\right)
    ,
    \mathbf{t}_{\mathrm{d},\mathit{i}}
\right), \label{eq:ld}
\end{equation}
\begin{equation}
    l_\mathrm{t}=l_\mathrm{gauss}\left(
    \mathbf{y}_\mathrm{t}\left(\mathbf{x}
    ,
    \mathbf{c}\right)
    ,
    \mathbf{t}_{\mathrm{t},\mathit{i}}
\right). \label{eq:lt}
\end{equation}
Here, $\mathbf{\Sigma}$ is originally a $3N \times 3N$ matrix; however, in this instance, it is assumed to have zeros in all the off-diagonal elements. The variance was predicted for each atom, resulting in different values for each atom; however, within the same atom, it was output with the same variance in the XYZ directions. By incorporating the standard deviation into the learning process, it is intended that the model will not need to fit outputs for structures that are difficult to learn, allowing it to better align the outputs for inputs that are easier to learn.
The fit of $\mathbf{y}_\mathrm{f}$ is determined using the cross-entropy error, represented by
\begin{equation}
l_{\mathrm{f}} =-\sum _{i_{\mathrm{feat}} =0}^{1}\log\frac{\exp([\mathbf{y}_{\mathrm{f}}(\mathbf{x},\mathbf{c})]_{i_{\mathrm{feat}}})}{\sum _{j_{\mathrm{feat}} =0}^{1}\exp([\mathbf{y}_{\mathrm{f}}(\mathbf{x} ,\mathbf{c})]_{j_{\mathrm{feat}}})}\mathrm{1}_{i_{\mathrm{feat}} =t_{\mathrm{f}}},
\label{eq:fin}
\end{equation}
where $[\mathbf{y}_{\mathrm{f},\mathit{i}}(\mathbf{x},\mathbf{c})]_{i_{\mathrm{feat}}}$ is the $i_{\mathrm{feat}}$-th element of the binary output $\mathbf{y}_{\mathrm{f},\mathit{i}}(\mathbf{x},\mathbf{c})$.
The loss during training was a linear combination of the transformation guidance field, denoising field, and stopping conditions. The coefficients for the loss related to $\mathbf{y}_{\mathrm{d}}$, the loss related to $\mathbf{y}_{\mathrm{t}}$, and the loss related to $\mathbf{y}_{\mathrm{f}}$ were all set to be the same.
The mean values of $\lVert\mathbf{t}_\mathrm{t}-\mathbf{y}_\mathrm{t}\rVert$ and $\lVert\mathbf{t}_\mathrm{d}-\mathbf{y}_\mathrm{d}\rVert$ for the training and validation data during the training steps are shown in Figure~\ref{fig:learning-curve-transition1x}. It can be observed that they decreased as the training progressed. Furthermore, while the loss of $\lVert\mathbf{t}_\mathrm{d}-\mathbf{y}_\mathrm{d}\rVert$ quickly decreases and then stops, $\lVert\mathbf{t}_\mathrm{t}-\mathbf{y}_\mathrm{t}\rVert$ continues to decrease for the training data. In the latter part, while $\lVert\mathbf{t}_\mathrm{t}-\mathbf{y}_\mathrm{t}\rVert$ continues to decrease for the training data, it ceases to decrease quickly for the validation data. This suggests that typical overfitting occurs. This overfitting implies that, even for datasets consisting of molecules with a similar number of atoms, the augmentation of reaction data could potentially contribute to performance improvement.

\section{Result}

\subsection{Definition of the field used in generation}
During the training steps, Eq.~\ref{eq:sample-train-data} was used. Similarly, using the outputs of the trained models $\mathbf{y}_\mathrm{t}, \mathbf{y}_\mathrm{d}$, the RP can be generated using
\begin{equation}
    \mathrm{d}\mathbf{x}=\mathbf{y}_\mathrm{t}(\mathbf{x},\mathbf{c})\mathrm{d}t+\alpha\mathbf{y}_\mathrm{d}(\mathbf{x},\mathbf{c})+g\mathrm{d}\mathbf{w} .\label{eq:simple-generation}
\end{equation}
When dealing with molecules that significantly exceed the size of the training data, there are cases in which the orientations of $\mathbf{y}_\mathrm{t}$ and $\mathbf{y}_\mathrm{d}$ are opposite. This can make generation difficult when using Equation~\ref{eq:simple-generation}. In such scenarios,
\begin{equation}
    \mathrm{d}\mathbf{x}=\mathbf{y}_{\mathrm{t}}(\mathbf{x},\mathbf{c})\mathrm{d}t+\alpha\mathbf{y}_{\mathrm{d}}^\top(\mathbf{x},\mathbf{c})+g\mathrm{d}\mathbf{w}
    \label{eq:perpendicular-generation}
\end{equation}
was used for the generation.
Here, $\mathbf{y}_{\mathrm{d}}^\top$ is an orthogonalized $\mathbf{y}_{\mathrm{d}}$ with respect to $\mathbf{y}_{\mathrm{t}}$ and defined as 
\begin{equation}
    \mathbf{y}_{\mathrm{d}}^\top(\mathbf{x},\mathbf{c})
    =
    \mathbf{y}_{\mathrm{d}}(\mathbf{x},\mathbf{c})
    -
    \frac{
        \mathbf{y}_{\mathrm{d}}(\mathbf{x},\mathbf{c})
        \cdot
        \mathbf{y}_{\mathrm{t}}(\mathbf{x},\mathbf{c})
    }{
        \mathbf{y}_{\mathrm{t}}(\mathbf{x},\mathbf{c})
        \cdot
        \mathbf{y}_{\mathrm{t}}(\mathbf{x},\mathbf{c})
    }
    \mathbf{y}_{\mathrm{t}}(\mathbf{x},\mathbf{c}).
    \label{eq:perpendicular-denoising}
\end{equation}
Additionally, for conditional generation, classifier-free guidance \cite{Ho2022-bw} as
\begin{equation}
    \mathbf{y}(\mathbf{x},\mathbf{c})=(1+w) \mathbf{y}(\mathbf{x},\mathbf{c})- w \mathbf{y}(\mathbf{x},\mathbf{0}) \label{eq:classifier-free-guidance}
\end{equation}
was used.
When utilizing classifier-free guidance for orthogonalization, Eq.~\ref{eq:perpendicular-denoising} was applied to both the conditional and unconditional outputs, followed by the application of Eq.~\ref{eq:classifier-free-guidance}. Afterward, orthogonalization was once again performed using Eq.~\ref{eq:perpendicular-denoising}.
Here, $\mathbf{y}(\mathbf{x},\mathbf{c})$ denotes the output vector under certain conditions and $\mathbf{y}(\mathbf{x},\mathbf{0})$ represents the output vector without conditions. In addition, the termination condition is determined using $\mathbf{y}_{\mathrm{f}}$. $\mathbf{y}_{\mathrm{f}}$ is a binary output, and the generation is terminated when the first element becomes larger than the zeroth element.

\subsection{The importance of the denoising field}
\begin{figure}
    \centering
    \includegraphics[width=1.0\linewidth]{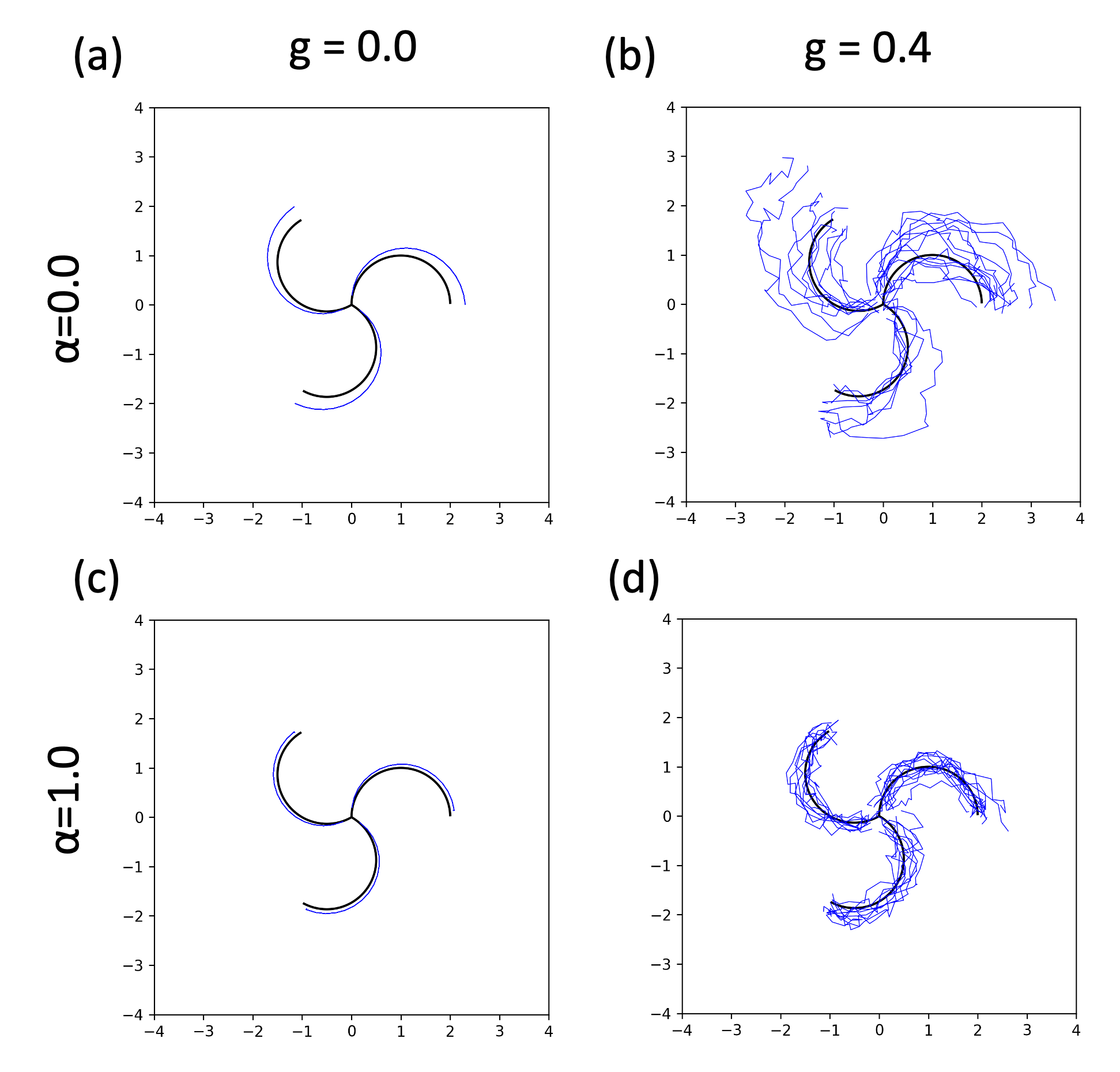}
   \caption{Results of generation with various parameters for the same RP as in Fig.~\ref{fig:flow}. $\alpha=0.0, g=0.0$, (b)$\alpha=0.0, g=0.4$, (c)$\alpha=1.0, g=0.0$, (d)$\alpha=1.0, g=0.4$}
    \label{fig:2dnoise}
\end{figure}

\begin{figure}
    \centering
    \includegraphics[width=1.0\linewidth]{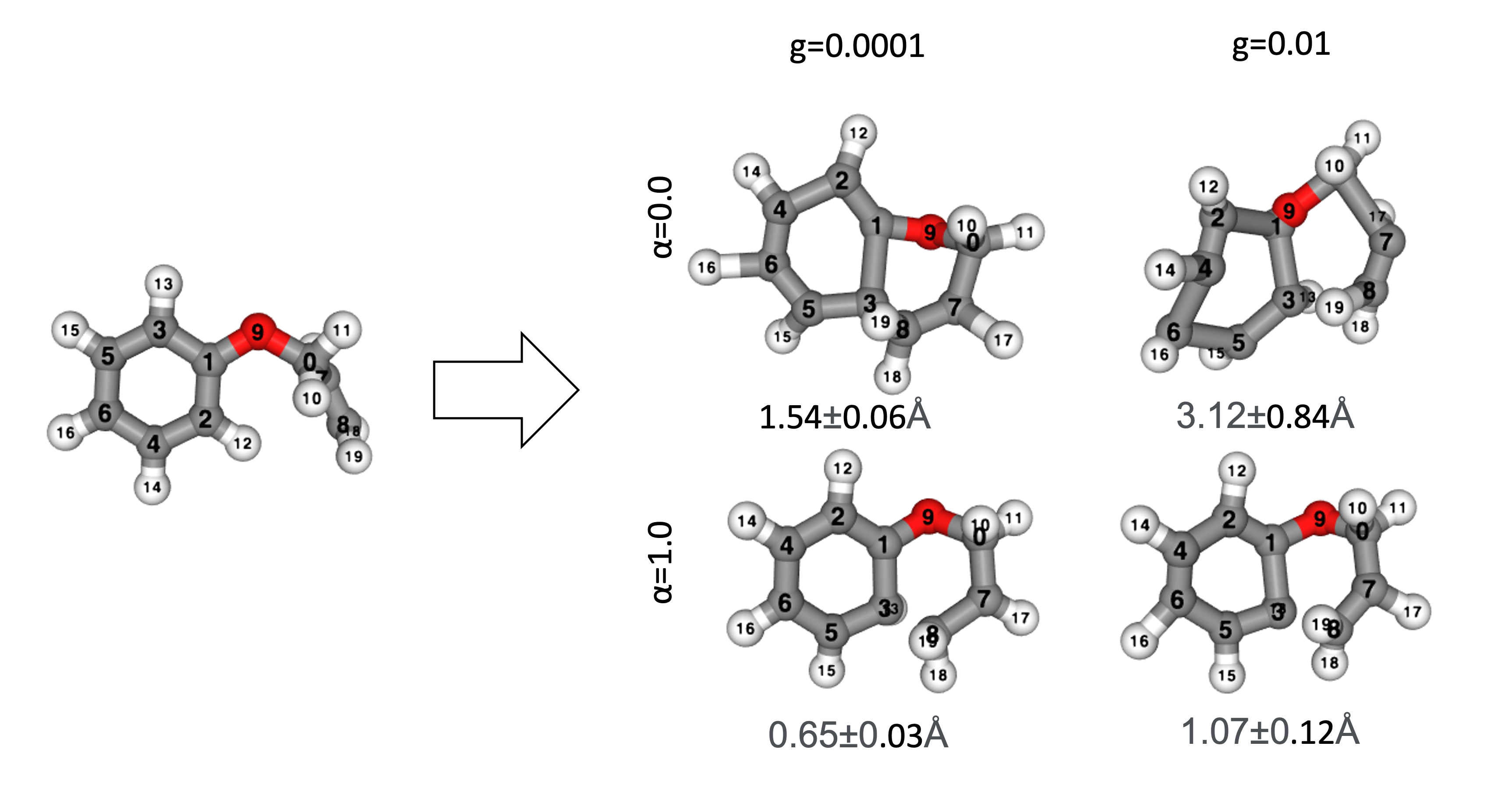}
    \caption{Results of the cyclization reaction of $\mathrm{C}_8\mathrm{H}_9\mathrm{O}$ generated with various parameters. (a) $\alpha=0.0, g=0.0001$, (b) $\alpha=0.0, g=0.01$, (c) $\alpha=1.0, g=0.0001$, (d) $\alpha=1.0, g=0.01$. The numbers written below each generation result represent the average and standard deviation of the RMSD between the original RP included in YARP and the optimized FS by PFP.}
    \label{fig:ndnoise}
\end{figure}

The generation was performed using denoising fields of various magnitudes to ascertain the importance of the denoising field. First, to verify the importance of the denoising field in a two-dimensional toy model, the same hypothetical RP shown in Fig.~\ref{fig:flow} was utilized, and generation was conducted using Eq.~\ref{eq:sample-train-data}. 
$\alpha\in\{0.0,1.0\}, g\in\{0.0, 0.4\}$ were used. The results are presented in Fig.~\ref{fig:2dnoise}. Even when $g=0.0$ where there is no noise, the path deviates from the true path when $\alpha=0.0$, whereas the deviation from the true path is minimal when $\alpha=1.0$. When $g=0.4$, the importance of the denoising field increases further; for $\alpha=0.0$, the path passing through points significantly deviates from the true path, whereas for $\alpha=1.0$, it shows a distribution encompassing the true path's vicinity.
Furthermore, actual molecular RPs were generated using various parameters. The IS and RP conditions were selected from one of the optimized $\mathrm{C_8H_9O}$ pathways and recalculated using PFP v4.0.0, found in the yarp dataset \cite{zhao2021yarp}. We performed the generation using $\alpha \in \{0.0, 1.0\}$ and $g \in \{0.0001, 0.01\}$. Generation was conducted 16 times for each parameter set, and the RMSD between the generated FS and the PFP optimized YARP FS was calculated. The average and standard deviations of the RMSD were calculated. The results are presented in Fig.~\ref{fig:ndnoise}. When the noise was large as $g=0.01$, the structure was almost completely distorted with $\alpha=0.0$, leaving very little of the original form. In contrast, at $\alpha=1.0, g=0.01$, although the structure was slightly degraded compared to the clean $\alpha=1.0, g=0.0001$ structure, it remained the same qualitatively. Moreover, even in the case of low noise, various bond angles are significantly distorted in the absence of a denoising field, as indicated by $\alpha=0.0, g=0.0001$.

\subsection{The limitation of the systematic generalization ability}
\begin{figure}
    \centering
    \includegraphics[width=1.0\linewidth]{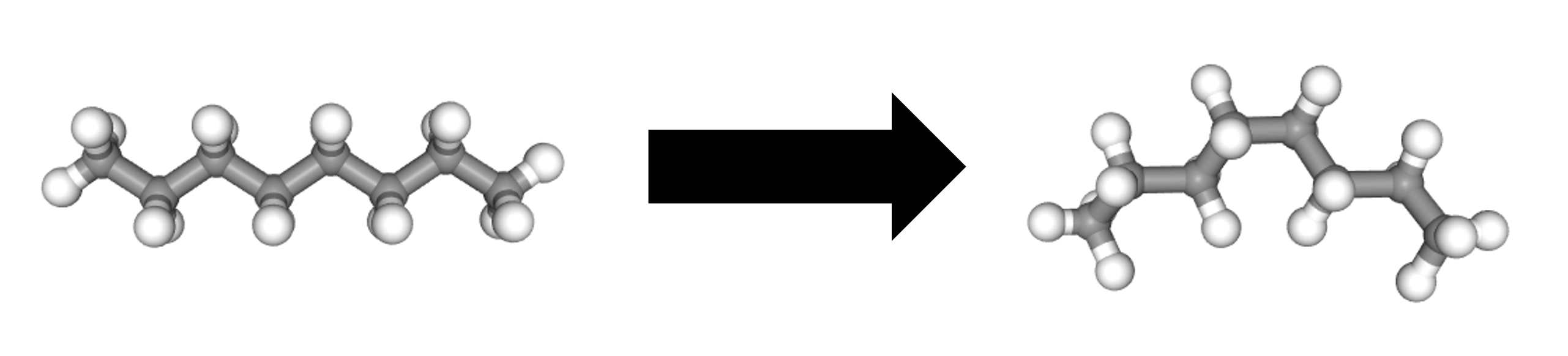}
    \caption{Example of the polyethylene rotational reaction used to test the limits of generalization performance.}
    \label{fig:polyethylene}
\end{figure}
The definitions of the denoising field, denoising coefficient, and size of the classifier-free guidance (Equation~\ref{eq:classifier-free-guidance}) were varied to generate the RPs for the dihedral angle rotation of the center of polyethylene of various lengths. For the generation, we used Eqs. ~\ref{eq:simple-generation} and ~\ref{eq:perpendicular-generation} with $\mathrm{d}t=0.1$. The value of $\alpha$ was chosen to be 1.0, the theoretical Newton step, and 0.1, which is smaller than 1.0.   For classifier-free guidance, we used $w\in\{0.0, 1.0, 2.0, 4.0, 8.0, 16.0\}$. For polyethylene with an even number of carbon atoms ($n$) ranging from 2 to 16, conditional generation was performed such that the dihedral angle of the central $\mathrm{CC}$-bond was rotated.
Generation success varies depending on the conditions, formulas, and parameters used. We first investigated whether the generation time was excessively long. Specifically, we verified whether the completion time was < 400 steps.

The reasons for the increased generation time are as follows.
(1), $\mathbf{y}_\mathrm{t}$ and $\mathbf{y}_{\mathrm{d}}$ are roughly oriented in opposite directions, causing the atoms to oscillate.
(2) After dissociation, the complex underwent configurational changes.
(3) The RP we attempted to generate was too long to represent within 400 steps.

(1) In such cases, the reaction does not proceed and often oscillates while maintaining a similar structure. However, there were instances in which the reaction progressed gradually while oscillating. (2) After the molecule splits, it moves without reaching an end. (3) This does not occur when trying to generate the rotation of the CC bond correctly but occurs when attempting to generate a much longer multistep RP.
The results are listed in Table~\ref{tbl:finally-finished}. It is particularly noticeable that many cases do not converge when generated using $\alpha = 1.0$ in Equation~\ref{eq:simple-generation}: Even in systems with sufficiently low carbon numbers where proper learning is expected, such as $n \le 7$ and the maximum $n$ for Transition1x, there were instances of non-convergence. Upon examination of these cases, it was found that in systems such as $n=2, w=1.0$ and $n=6, w=16.0$, unintended hydrogen migration occurred. The convergence criteria were not met even after the migration, leading to sustained oscillations. For $n=6, w=1.0$ and $n=8, w=8.0, 2.0, 1.0, 0.0$, the reaction did not proceed, and the structure oscillated around the IS.
Even when Equation~\ref{eq:simple-generation}, convergence tends to be easier when $\alpha=0.1$, although convergence was often not achieved for $n=16$ in these instances, many other cases resulted in convergence.

One possible reason for the reaction remaining in the IS without any progress is that $\mathbf{y}_\mathrm{t}$ and $\mathbf{y}_\mathrm{d}$ may indicate opposite directions. Therefore, the generation was performed using Equation~\ref{eq:perpendicular-generation}. The results of generation using Equation~\ref{eq:perpendicular-generation} are listed in the lower section of Table~\ref{tbl:finally-finished}. It was observed that, even with $\alpha=1.0$, many examples satisfied the convergence conditions. The number of examples converged for $\alpha=0.1$ was greater than for Equation~\ref{eq:simple-generation}.

In the data of Table  \ref{tbl:finally-finished}, it was only verified whether the machine learning model has determined that the generation should be stopped, and the validity of the generated results was not confirmed. Therefore, we investigated whether the obtained results satisfied the specified FS. We summarized them in Tables ~\ref{tbl:finally-satisfied-simple-0.1} and ~\ref{tbl:always-satisfied-simple-0.1}. Table~\ref{tbl:finally-satisfied-simple-0.1} summarizes whether the FS has a central $\mathrm{CC}$ bond rotated compared to the IS and whether other bonds are the same as the IS. Table~\ref{tbl:always-satisfied-simple-0.1} includes only those cases in which the angles of unrelated bonds do not vary significantly throughout the RP, as indicated by a check mark. When $w=0.0$ without classifier-free guidance, the specified conditions were ignored for all cases. The reaction was derived without central $\mathrm{CC}$ bond rotation. When $\alpha=0.1$ is used in Equation~\ref{eq:simple-generation}, an easily understandable result is obtained, where the larger the value of $w$, the more likely the conditions are to be satisfied. Moreover, the FS satisfied the given conditions up to $n=12$, which exceeded the training data size. However, when using $\alpha=1.0$ in Equation~\ref{eq:simple-generation}, there are very few examples in which generation satisfying the conditions is performed. Using Equation~\ref{eq:perpendicular-generation}, more examples satisfied the conditions than using Equation~\ref{eq:simple-generation} even when $\alpha=1.0$ was applied.
In summary, employing Equation~\ref{eq:simple-generation} with $\alpha=0.1$ yields manageable conditional generation results for small molecules. Conversely, when deploying a trained model in practical applications, the generation should be completed in a finite number of steps, even for larger datasets that exceed the training data size. For this purpose, Equation~\ref{eq:perpendicular-generation} can be used.
\begin{table}
\caption{When generated under various conditions, whether the termination condition was met within 400 steps. The parts with check marks indicate cases where the termination condition was met within 400 steps. In each table, the horizontal axis ($w$) indicates the strength of the classifier-free guidance (Eq.~\ref{eq:classifier-free-guidance}). The vertical axis ($n$) indicates the number of carbons in polyethylene. The tables at the top left and top right are generated using Eq.~\ref{eq:simple-generation}. The tables at the bottom left and bottom right are generated using Eq.~\ref{eq:perpendicular-generation}. The tables on the top left and bottom left used $\alpha=0.1$. The tables on the top right and bottom right used $\alpha=1.0$. Here, $\alpha$ is the coefficient for denoising.}
\centering
\label{tbl:finally-finished}
    \begin{tabular}{ccc}
        & $\alpha=0.1$ & $\alpha=1.0$ \\
        \rotatebox{90}{Eq.~\ref{eq:simple-generation}} & 
\begin{tabular}{lllllll}
\toprule
$n\backslash w$ & 0 & 1 & 2 & 4 & 8 & 16 \\
\midrule
2 & $\checkmark$ & $\checkmark$ & $\checkmark$ & $\checkmark$ & $\checkmark$ & $\checkmark$ \\
4 & $\checkmark$ & $\checkmark$ & $\checkmark$ & $\checkmark$ & $\checkmark$ & $\checkmark$ \\
6 & $\checkmark$ & $\checkmark$ & $\quad$ & $\checkmark$ & $\checkmark$ & $\checkmark$ \\
8 & $\checkmark$ & $\checkmark$ & $\checkmark$ & $\checkmark$ & $\checkmark$ & $\checkmark$ \\
10 & $\checkmark$ & $\checkmark$ & $\checkmark$ & $\checkmark$ & $\checkmark$ & $\checkmark$ \\
12 & $\checkmark$ & $\checkmark$ & $\checkmark$ & $\checkmark$ & $\checkmark$ & $\quad$ \\
14 & $\checkmark$ & $\checkmark$ & $\quad$ & $\checkmark$ & $\checkmark$ & $\checkmark$ \\
16 & $\quad$ & $\quad$ & $\quad$ & $\quad$ & $\checkmark$ & $\checkmark$ \\
\bottomrule
\end{tabular}
        &
\begin{tabular}{lllllll}
\toprule
$n\backslash w$ & 0 & 1 & 2 & 4 & 8 & 16 \\
\midrule
2 & $\checkmark$ & $\quad$ & $\checkmark$ & $\checkmark$ & $\checkmark$ & $\checkmark$ \\
4 & $\checkmark$ & $\checkmark$ & $\checkmark$ & $\checkmark$ & $\checkmark$ & $\checkmark$ \\
6 & $\checkmark$ & $\quad$ & $\checkmark$ & $\checkmark$ & $\checkmark$ & $\quad$ \\
8 & $\quad$ & $\quad$ & $\quad$ & $\checkmark$ & $\quad$ & $\checkmark$ \\
10 & $\quad$ & $\quad$ & $\quad$ & $\checkmark$ & $\checkmark$ & $\checkmark$ \\
12 & $\checkmark$ & $\quad$ & $\quad$ & $\quad$ & $\quad$ & $\checkmark$ \\
14 & $\checkmark$ & $\checkmark$ & $\checkmark$ & $\quad$ & $\quad$ & $\checkmark$ \\
16 & $\quad$ & $\quad$ & $\quad$ & $\quad$ & $\quad$ & $\checkmark$ \\
\bottomrule
\end{tabular}
        \\
        \rotatebox{90}{Eq.~\ref{eq:perpendicular-generation}} & 
\begin{tabular}{lllllll}
\toprule
$n\backslash w$ & 0 & 1 & 2 & 4 & 8 & 16 \\
\midrule
2 & $\checkmark$ & $\checkmark$ & $\checkmark$ & $\checkmark$ & $\checkmark$ & $\quad$ \\
4 & $\checkmark$ & $\checkmark$ & $\checkmark$ & $\checkmark$ & $\checkmark$ & $\checkmark$ \\
6 & $\checkmark$ & $\checkmark$ & $\checkmark$ & $\checkmark$ & $\quad$ & $\checkmark$ \\
8 & $\checkmark$ & $\checkmark$ & $\checkmark$ & $\checkmark$ & $\checkmark$ & $\checkmark$ \\
10 & $\checkmark$ & $\checkmark$ & $\checkmark$ & $\checkmark$ & $\checkmark$ & $\checkmark$ \\
12 & $\checkmark$ & $\checkmark$ & $\checkmark$ & $\checkmark$ & $\checkmark$ & $\checkmark$ \\
14 & $\quad$ & $\quad$ & $\checkmark$ & $\checkmark$ & $\checkmark$ & $\checkmark$ \\
16 & $\checkmark$ & $\checkmark$ & $\quad$ & $\checkmark$ & $\checkmark$ & $\checkmark$ \\
\bottomrule
\end{tabular}
        &
\begin{tabular}{lllllll}
\toprule
$n\backslash w$ & 0 & 1 & 2 & 4 & 8 & 16 \\
\midrule
2 & $\checkmark$ & $\checkmark$ & $\checkmark$ & $\checkmark$ & $\quad$ & $\checkmark$ \\
4 & $\checkmark$ & $\checkmark$ & $\checkmark$ & $\checkmark$ & $\quad$ & $\checkmark$ \\
6 & $\checkmark$ & $\checkmark$ & $\checkmark$ & $\checkmark$ & $\checkmark$ & $\checkmark$ \\
8 & $\checkmark$ & $\checkmark$ & $\checkmark$ & $\checkmark$ & $\checkmark$ & $\checkmark$ \\
10 & $\checkmark$ & $\checkmark$ & $\checkmark$ & $\checkmark$ & $\checkmark$ & $\checkmark$ \\
12 & $\checkmark$ & $\checkmark$ & $\checkmark$ & $\checkmark$ & $\checkmark$ & $\checkmark$ \\
14 & $\checkmark$ & $\checkmark$ & $\checkmark$ & $\checkmark$ & $\checkmark$ & $\checkmark$ \\
16 & $\quad$ & $\checkmark$ & $\checkmark$ & $\checkmark$ & $\checkmark$ & $\checkmark$ \\
\bottomrule
\end{tabular}
    \end{tabular}
\end{table}

\begin{table}
\caption{When generating under various conditions, the final product was checked for correctness to see if it matched the specified expectations. The parts marked with a check mark indicate correctly generated sets. In each table, the horizontal axis ($w$) represents the strength of classifier-free guidance (Eq.~\ref{eq:classifier-free-guidance}). The vertical axis ($n$) indicates the number of carbons in polyethylene. The tables in the upper left and upper right were generated using Eq.~\ref{eq:simple-generation}. The tables in the lower left and lower right were generated using Eq.~\ref{eq:perpendicular-generation}. The tables in the upper left and lower left utilized $\alpha=0.1$. The tables in the upper right and lower right utilized $\alpha=1.0$. Here, $\alpha$ represents the coefficient of denoising.}
\centering
\label{tbl:finally-satisfied-simple-0.1}
    \begin{tabular}{ccc}
        & $\alpha=0.1$ & $\alpha=1.0$ \\
        \rotatebox{90}{Eq.~\ref{eq:simple-generation}} & 
        \begin{tabular}{lllllll}
            \toprule
            $n\backslash w$ & 0 & 1 & 2 & 4 & 8 & 16 \\
            \midrule
            2 & $\quad$ & $\checkmark$ & $\checkmark$ & $\checkmark$ & $\checkmark$ & $\checkmark$ \\
            4 & $\quad$ & $\checkmark$ & $\checkmark$ & $\checkmark$ & $\checkmark$ & $\checkmark$ \\
            6 & $\quad$ & $\quad$ & $\quad$ & $\quad$ & $\checkmark$ & $\checkmark$ \\
            8 & $\quad$ & $\quad$ & $\quad$ & $\quad$ & $\checkmark$ & $\checkmark$ \\
            10 & $\quad$ & $\quad$ & $\checkmark$ & $\quad$ & $\checkmark$ & $\quad$ \\
            12 & $\quad$ & $\quad$ & $\quad$ & $\quad$ & $\checkmark$ & $\checkmark$ \\
            14 & $\quad$ & $\quad$ & $\quad$ & $\quad$ & $\quad$ & $\quad$ \\
            16 & $\quad$ & $\quad$ & $\quad$ & $\quad$ & $\quad$ & $\quad$ \\
            \bottomrule
        \end{tabular}
        &
        \begin{tabular}{lllllll}
            \toprule
            $n\backslash w$ & 0 & 1 & 2 & 4 & 8 & 16 \\
            \midrule
            2 & $\quad$ & $\quad$ & $\quad$ & $\checkmark$ & $\quad$ & $\quad$ \\
            4 & $\quad$ & $\quad$ & $\quad$ & $\quad$ & $\quad$ & $\quad$ \\
            6 & $\quad$ & $\quad$ & $\quad$ & $\checkmark$ & $\quad$ & $\quad$ \\
            8 & $\quad$ & $\quad$ & $\quad$ & $\checkmark$ & $\quad$ & $\quad$ \\
            10 & $\quad$ & $\quad$ & $\quad$ & $\quad$ & $\checkmark$ & $\quad$ \\
            12 & $\quad$ & $\quad$ & $\quad$ & $\quad$ & $\quad$ & $\quad$ \\
            14 & $\quad$ & $\quad$ & $\quad$ & $\quad$ & $\quad$ & $\quad$ \\
            16 & $\quad$ & $\quad$ & $\quad$ & $\quad$ & $\quad$ & $\quad$ \\
            \bottomrule
        \end{tabular}
        \\
        \rotatebox{90}{Eq.~\ref{eq:perpendicular-generation}} & 
        \begin{tabular}{lllllll}
            \toprule
            $n\backslash w$ & 0 & 1 & 2 & 4 & 8 & 16 \\
            \midrule
            2 & $\quad$ & $\checkmark$ & $\checkmark$ & $\checkmark$ & $\checkmark$ & $\quad$ \\
            4 & $\quad$ & $\checkmark$ & $\checkmark$ & $\checkmark$ & $\checkmark$ & $\checkmark$ \\
            6 & $\quad$ & $\quad$ & $\quad$ & $\checkmark$ & $\quad$ & $\checkmark$ \\
            8 & $\quad$ & $\quad$ & $\quad$ & $\checkmark$ & $\quad$ & $\checkmark$ \\
            10 & $\quad$ & $\quad$ & $\quad$ & $\quad$ & $\checkmark$ & $\checkmark$ \\
            12 & $\quad$ & $\quad$ & $\quad$ & $\quad$ & $\quad$ & $\quad$ \\
            14 & $\quad$ & $\quad$ & $\quad$ & $\quad$ & $\quad$ & $\quad$ \\
            16 & $\quad$ & $\quad$ & $\quad$ & $\quad$ & $\quad$ & $\quad$ \\
            \bottomrule
        \end{tabular}
        &
        \begin{tabular}{lllllll}
            \toprule
            $n\backslash w$ & 0 & 1 & 2 & 4 & 8 & 16 \\
            \midrule
            2 & $\quad$ & $\checkmark$ & $\checkmark$ & $\checkmark$ & $\quad$ & $\quad$ \\
            4 & $\quad$ & $\quad$ & $\checkmark$ & $\checkmark$ & $\quad$ & $\quad$ \\
            6 & $\checkmark$ & $\quad$ & $\checkmark$ & $\quad$ & $\quad$ & $\checkmark$ \\
            8 & $\quad$ & $\quad$ & $\quad$ & $\quad$ & $\checkmark$ & $\quad$ \\
            10 & $\quad$ & $\quad$ & $\quad$ & $\quad$ & $\checkmark$ & $\quad$ \\
            12 & $\quad$ & $\quad$ & $\quad$ & $\quad$ & $\quad$ & $\quad$ \\
            14 & $\quad$ & $\quad$ & $\quad$ & $\quad$ & $\quad$ & $\quad$ \\
            16 & $\quad$ & $\quad$ & $\quad$ & $\quad$ & $\quad$ & $\quad$ \\
            \bottomrule
        \end{tabular}
    \end{tabular}
\end{table}

\begin{table}
\caption{When generated under various conditions, the final product was checked to determine whether it was as specified, did not include rotations other than the specified bonds during the process, and whether there were no changes in the bonds during the process. The parts with check marks indicate correctly generated sets. In each table, the horizontal axis ($w$) indicates the strength of classifier-free guidance (Eq.~\ref{eq:classifier-free-guidance}). The vertical axis ($n$) indicates the number of carbons in polyethylene. The top left and top right tables were generated using Eq.~\ref{eq:simple-generation}. The bottom left and bottom right tables were generated using Eq.~\ref{eq:perpendicular-generation}. The top left and bottom left tables used $\alpha=0.1$, and the top right and bottom right tables used $\alpha=1.0$. Here, $\alpha$ is the coefficient for denoising.}
\centering
\label{tbl:always-satisfied-simple-0.1}
    \begin{tabular}{ccc}
        & $\alpha=0.1$ & $\alpha=1.0$ \\
        \rotatebox{90}{Eq.~\ref{eq:simple-generation}} & 
\begin{tabular}{lllllll}
\toprule
$n\backslash w$ & 0 & 1 & 2 & 4 & 8 & 16 \\
\midrule
2 & $\quad$ & $\checkmark$ & $\checkmark$ & $\checkmark$ & $\checkmark$ & $\checkmark$ \\
4 & $\quad$ & $\checkmark$ & $\checkmark$ & $\checkmark$ & $\checkmark$ & $\checkmark$ \\
6 & $\quad$ & $\quad$ & $\quad$ & $\quad$ & $\checkmark$ & $\checkmark$ \\
8 & $\quad$ & $\quad$ & $\quad$ & $\quad$ & $\checkmark$ & $\checkmark$ \\
10 & $\quad$ & $\quad$ & $\quad$ & $\quad$ & $\quad$ & $\quad$ \\
12 & $\quad$ & $\quad$ & $\quad$ & $\quad$ & $\quad$ & $\quad$ \\
14 & $\quad$ & $\quad$ & $\quad$ & $\quad$ & $\quad$ & $\quad$ \\
16 & $\quad$ & $\quad$ & $\quad$ & $\quad$ & $\quad$ & $\quad$ \\
\bottomrule
\end{tabular}
        &
\begin{tabular}{lllllll}
\toprule
$n\backslash w$ & 0 & 1 & 2 & 4 & 8 & 16 \\
\midrule
2 & $\quad$ & $\quad$ & $\quad$ & $\checkmark$ & $\quad$ & $\quad$ \\
4 & $\quad$ & $\quad$ & $\quad$ & $\quad$ & $\quad$ & $\quad$ \\
6 & $\quad$ & $\quad$ & $\quad$ & $\checkmark$ & $\quad$ & $\quad$ \\
8 & $\quad$ & $\quad$ & $\quad$ & $\checkmark$ & $\quad$ & $\quad$ \\
10 & $\quad$ & $\quad$ & $\quad$ & $\quad$ & $\quad$ & $\quad$ \\
12 & $\quad$ & $\quad$ & $\quad$ & $\quad$ & $\quad$ & $\quad$ \\
14 & $\quad$ & $\quad$ & $\quad$ & $\quad$ & $\quad$ & $\quad$ \\
16 & $\quad$ & $\quad$ & $\quad$ & $\quad$ & $\quad$ & $\quad$ \\
\bottomrule
\end{tabular}
        \\
        \rotatebox{90}{Eq.~\ref{eq:perpendicular-generation}} & 
\begin{tabular}{lllllll}
\toprule
$n\backslash w$ & 0 & 1 & 2 & 4 & 8 & 16 \\
\midrule
2 & $\quad$ & $\checkmark$ & $\checkmark$ & $\checkmark$ & $\checkmark$ & $\quad$ \\
4 & $\quad$ & $\checkmark$ & $\checkmark$ & $\checkmark$ & $\checkmark$ & $\checkmark$ \\
6 & $\quad$ & $\quad$ & $\quad$ & $\checkmark$ & $\quad$ & $\checkmark$ \\
8 & $\quad$ & $\quad$ & $\quad$ & $\checkmark$ & $\quad$ & $\checkmark$ \\
10 & $\quad$ & $\quad$ & $\quad$ & $\quad$ & $\quad$ & $\quad$ \\
12 & $\quad$ & $\quad$ & $\quad$ & $\quad$ & $\quad$ & $\quad$ \\
14 & $\quad$ & $\quad$ & $\quad$ & $\quad$ & $\quad$ & $\quad$ \\
16 & $\quad$ & $\quad$ & $\quad$ & $\quad$ & $\quad$ & $\quad$ \\
\bottomrule
\end{tabular}
        &
\begin{tabular}{lllllll}
\toprule
$n\backslash w$ & 0 & 1 & 2 & 4 & 8 & 16 \\
\midrule
2 & $\quad$ & $\checkmark$ & $\checkmark$ & $\checkmark$ & $\quad$ & $\quad$ \\
4 & $\quad$ & $\quad$ & $\checkmark$ & $\checkmark$ & $\quad$ & $\quad$ \\
6 & $\checkmark$ & $\quad$ & $\checkmark$ & $\quad$ & $\quad$ & $\checkmark$ \\
8 & $\quad$ & $\quad$ & $\quad$ & $\quad$ & $\checkmark$ & $\quad$ \\
10 & $\quad$ & $\quad$ & $\quad$ & $\quad$ & $\checkmark$ & $\quad$ \\
12 & $\quad$ & $\quad$ & $\quad$ & $\quad$ & $\quad$ & $\quad$ \\
14 & $\quad$ & $\quad$ & $\quad$ & $\quad$ & $\quad$ & $\quad$ \\
16 & $\quad$ & $\quad$ & $\quad$ & $\quad$ & $\quad$ & $\quad$ \\
\bottomrule
\end{tabular}
    \end{tabular}
\end{table}

\subsection{Generation examples}\label{sec:examples}
\begin{figure}
    \centering
    \includegraphics[width=1\linewidth]{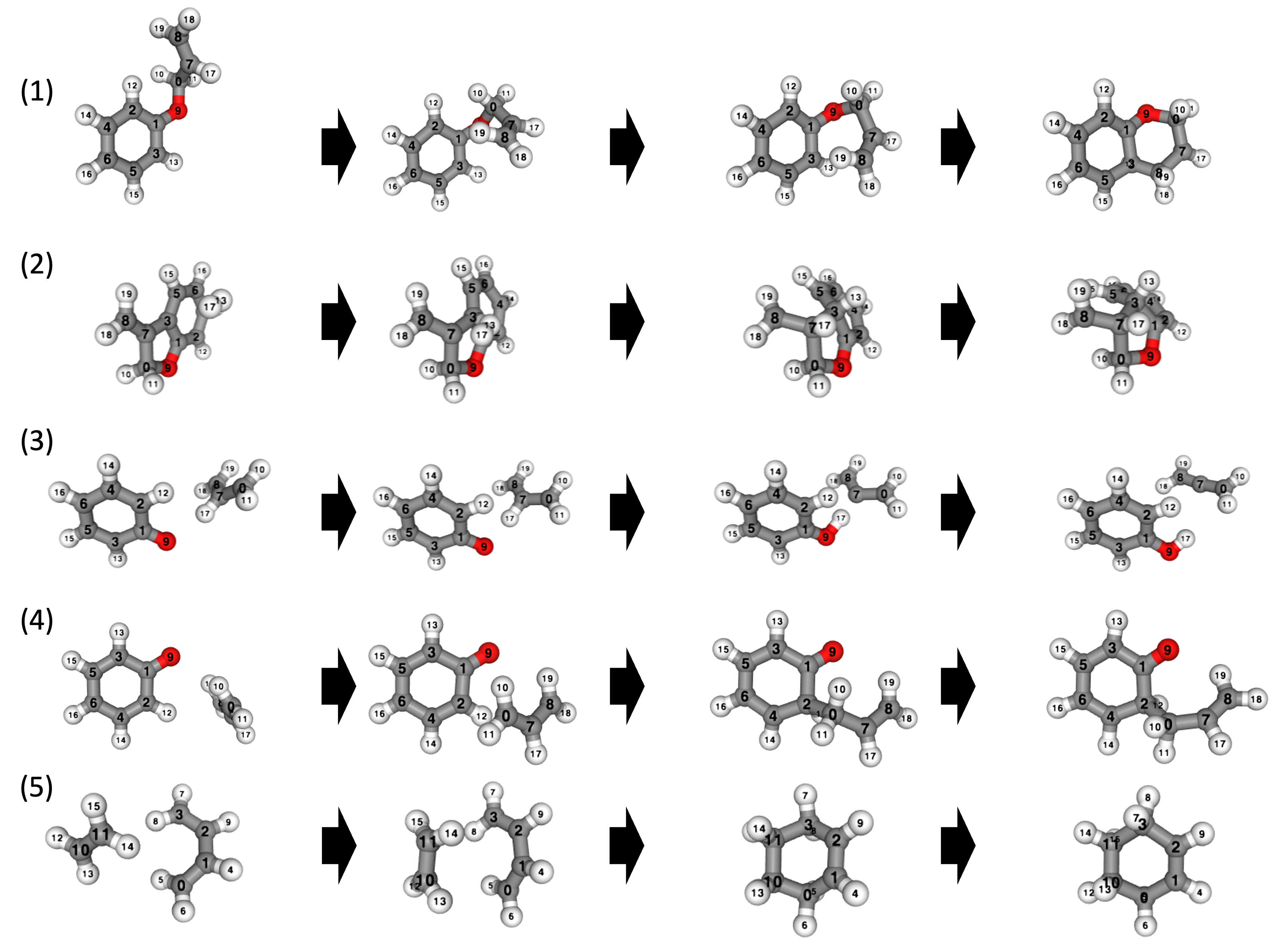}
    \caption{Results of conditional production for ${\rm C_8H_9O}$ and Diels-Alder reaction. (1) We set the condition $c$ to bond the 3rd atom and the 8th atom. (2) We set the condition $c$ to bond the 7th atom and the 17th atom, and bond the 0th atom and the 11th atom, while dissociating the 17th atom and the 11th atom. (3) We set the condition $c$ to separate the 17th atom and the 7th atom, and bond the 17th atom and the 9th atom. (4) We set the condition $c$ to bond the 0th atom and the 2nd atom. (5) We set the condition $c$ to bond the 11th atom and the 3rd atom, and bond the 10th atom and the 3rd atom.}
    \label{fig:examples}
\end{figure}

Examples of the results of the conditional generation given the changes in bonding are shown in Figure~\ref{fig:examples}. Generation is performed using Eq.~\ref{eq:simple-generation}, which does not involve orthogonalization with $\alpha=0.1$. Moreover, classifier-free guidance with $w=4$ is used (Equation~\ref{eq:classifier-free-guidance}). The chosen molecules were reactions involving $\mathrm{C}_8\mathrm{H}_9\mathrm{O}$ from the yarp dataset \cite{zhao2021yarp} and a typical Diels-Alder reaction. The training data used were Transition1x, which only included up to $\mathrm{C}_7$; however, despite this, the trained model was still able to generate reactions for this molecule, $\mathrm{C}_8\mathrm{H}_9\mathrm{O}$.

The first reaction in Figure~\ref{fig:examples} involves rotation of the dihedral angles and formation of bonds. Manually creating such an initial RP without the use of a neural network requires meticulous manipulation of the dihedral angle rotations and bond formations, making it an extremely labor-intensive task. This neural network can generate this reaction within seconds by simply specifying the bonds.
Additionally, the first reaction in Figure~\ref{fig:examples} includes dihedral rotation, and although the reaction coordinate is significantly curved from the Cartesian coordinates, it can still provide a qualitatively correct RP. The other reactions shown in Figure~\ref{fig:examples} encompass various reactions. The second reaction involves the addition of a hydrogen atom to the double bond. The third reaction is a hydrogen transfer reaction. The fourth is a bimolecular coupling reaction. The fifth reaction is the Diels-Alder reaction. These reactions can also be generated qualitatively and correctly.

\subsection{Conditional generation for various types of data}\label{sec:conditional}
\begin{figure}
    \centering
    \includegraphics[width=1\linewidth]{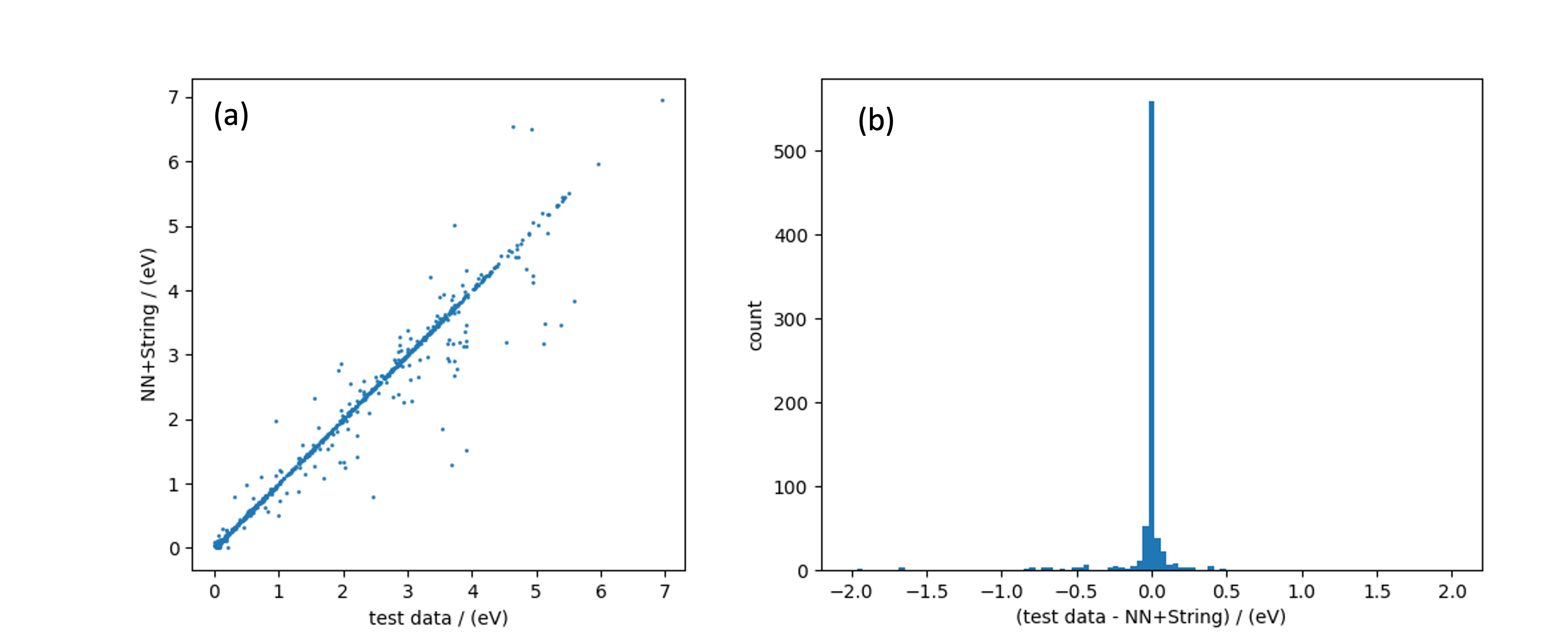}
    \caption{(a) Comparison of activation barriers for reactions in test data set vs. activation barriers for generated-optimized results. (b) The histogram for energy difference between test data and generated-optimized RP.}
    \label{fig:compare-generation}
\end{figure}

Conditional generation was performed on 833 test data points separated from training data. Eq.~\ref{eq:simple-generation}, which does not involve orthogonalization with $\alpha=0.1$ was used for the conditional generation, and Eq.~\ref{eq:classifier-free-guidance} was employed. As the value of \( w \) increases, reactions fulfilling the given conditions proceed with more data. The results are presented in Table.~\ref{tab:condition-succeed}.
\begin{table}
    \centering
    \begin{tabular}{ccc}\toprule
            $w$&  number& ratio\\
            \midrule
            0.0&  519& 0.62\\
 0.5& 662&0.79\\
            1.0&  714& 0.85\\
 2.0& 753&0.90\\
            4.0&  785& 0.94\\
 8.0& 790&0.95\\
 \bottomrule
    \end{tabular}
    \caption{\(w\) is the parameter for classifier-free guidance. The column "number" is the number of generative results that satisfy the generation conditions of the RPs. The column "ratio" is the number of successful pathways divided by the number of test data.}
    \label{tab:condition-succeed}
\end{table}
Among the results obtained using classifier-free guidance with $w=8.0$, those satisfying the given conditions were optimized using the string method at the PFP v4.0.0 level. The activation barriers of the test datasets were compared. In this comparison, we expected the same activation barrier to be obtained if the output followed the same qualitative pathway. The results are presented in Figure \ref{fig:compare-generation}(a). A histogram of the energy differences between the generated and optimized RP and the test RP is shown in Figure \ref{fig:compare-generation}(b). Most reactions were expected to follow a qualitatively similar RP and demonstrate the same activation barriers as the test dataset.
However, some reactions followed pathways completely different from those in the test dataset and yielded qualitatively different activation barriers. Generally, there are qualitatively different RPs, even if the bonding changes are the same. In this experiment, there were instances in which reactions with activation barriers higher than the test data were obtained; conversely, pathways with barriers lower than the test data were also found. Therefore, we believe that the model obtained in this study is well-learned as far as this experiment can confirm.

\subsection{Random generation}
\begin{figure}
    \centering
    \includegraphics[width=1.0\linewidth]{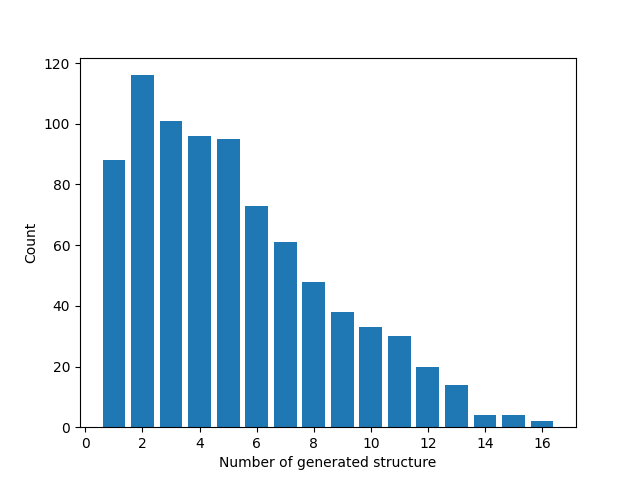}
    \caption{Plot of the types of FS structures obtained after 16 random generations from an IS. The horizontal axis represents the number of the types of FS structures obtained. The vertical axis represents the number of ISs that yielded those FS structures.}
    \label{fig:result-random}
\end{figure}

White noise was added during the generation. Equation \ref{eq:simple-generation} was used for this generation. Structural optimization was performed before and after generation, and changes in the bonds and rotations were analyzed. The number of times the generation resulted in different changes from the same IS was counted. Here, 16 was arbitrarily selected as the artificial cutoff. Because the number of test datasets was large, each reaction was generated 16 times to reduce the computational cost of the experiment. 0.1 was used for $\mathrm{d}t$, 0.1 for $g$, and 0.1 for $\alpha$. A histogram of the results is shown in Figure~\ref{fig:result-random}. The most frequent values on the horizontal axis were 1 and 2, which indicates that many cases of similar reactions were obtained no matter how many times they were generated. However, there are some examples where the value on the horizontal axis is significantly higher. Different reactions can be achieved using the same IS.

\section{Conclusion}
A machine learning model for reaction-path generation was proposed. The model can obtain an approximate sketch of the entire reaction pathway with several Neural Network evaluations. The model can handle reaction pathways in $3N$-dimensional space and generate complex reactions such as chemical reactions in organic chemistry. The model could learn and generalize the transition1x results to generate reactions similar to those of the test set. This model can be used for both conditional and random generations. In this experiment, conditional generation was performed using changes in the bonding. Response paths that accurately reproduced the bond changes in the test data were generated with the proper use of classifier-free guidance. In addition, when randomness was included in the generation, many reactions were generated in each initial state.

\section{Acknowledgement}
We would like to extend our heartfelt gratitude to the members of Preferred Networks, Inc., with special thanks to Kohei Shinohara, Yunzhuo Wang, Hirotaka Akita, Hodaka Mori, Iori Kurata, Yu Miyazaki, and the entire team for their invaluable comments, discussions, and unwavering support.

\bibliographystyle{achemso}
\bibliography{references,referencesadditional} 

\end{document}